\documentclass[aps,prev,onecolumn,preprintnumbers,aps,
showkeys,floatfix,nofootinbib,brazil]{revtex4-1}

\usepackage[utf8]{inputenc}
\usepackage{lmodern}			
\usepackage[T1]{fontenc}		
\usepackage[brazil]{babel}	
\usepackage{graphicx}
\usepackage{color}
\usepackage{dcolumn}
\usepackage{bm}
\usepackage{epsf}
\usepackage{mathtools}
\usepackage{mathptmx}
\usepackage{yhmath}
\usepackage{amsmath}
\usepackage{natbib}
\usepackage{eucal}
\graphicspath{{figArticle/}}
\usepackage{float}

\usepackage{multirow}

\begin{document}
\title{Constraints on Cosmographic Functions of Cosmic Chronometers Data Using Gaussian Processes}

\author{A. M. Velasquez-Toribio}
\email{alan.toribio@ufes.br}
\affiliation{Nucleo Cosmo-ufes \& Departamento de F\'{\i}sica, Universidade Federal do Espirito Santo,  29075-910 Vit\'{\o}ria - ES, Brasil}

\author{Júlio C. Fabris}
\email{julio.fabris@cosmo-ufes.org}
\affiliation{Nucleo Cosmo-ufes \& Departamento de F\'{\i}sica, Universidade Federal do Espirito Santo,  29075-910 Vit\'{\o}ria - ES, Brasil} \affiliation{National  Research  Nuclear  University  MEPhI,  Kashirskoe  sh.   31,  Moscow  115409, Russia}

\date{\today}

\begin{abstract}
We study observational constraints on the cosmographic 
functions up to the fourth derivative of the scale factor 
with respect to cosmic time, i.e., the so-called snap function, using
the non-parametric method of Gaussian Processes.
As observational data we use the Hubble parameter data.
Also we use mock data sets to estimate the future forecast 
and study the performance of this type of data to constrain
cosmographic functions. The combination between a non-parametric 
method and the Hubble parameter data is investigated as a strategy to reconstruct 
cosmographic functions.
In addition, our results are quite general because 
they are not restricted to a specific type of functional dependency of the Hubble parameter.
We investigate some advantages of using cosmographic functions instead of cosmographic series, 
since the former are general definitions free of approximations. 
In general, our results do not deviate significantly from $\Lambda CDM$. 
We determine a transition redshift $z_{tr}=0.637^{+0.165}_{-0.175}$ and $H_{0}=69.45 \pm 4.34$. Also assuming priors
for the Hubble constant we obtain
$z_{tr}=0.670^{+0.210}_{-0.120}$ with $H_{0}=67.44$ (Planck) and $z_{tr}=0.710^{+0.159}_{-0.111}$ with $H_{0}=74.03$(SH0ES).
Our main results are summarized in table 2.

\end{abstract}
\pacs{98.80.-k, 95.36.+x, 98.80.Es}
\maketitle

\section{Introduction}

The accelerated expansion of the Universe \cite{riess1998, perlmutter1998} 
is currently one of the main problems of cosmology.
The simplest explantion is given by the $\Lambda CDM$ model 
making use of cold dark matter and a cosmological constant
which matches well with different observational data types:
supernovae Ia, cosmic microwave background (CMB), baryons acoustic ocillations (BAO), 
etc, see e.g. \cite{scolnic, planck, Baosdss}.

The $\Lambda CDM$ model includes a component of dark energy which is 
responsible for the accelerating expansion of the universe.
However, from a theoretical point of view, one of the big problems that affect this 
model is the difference between 
the measured value of the cosmological constant and that required by quantum theory 
(depending on the energy scale), which can reach a part in $10^{120}$. 
This gives rise to the cosmological constant fine tuning problem. Other theoretical 
problem is the cosmological coincidence problem, i.e., 
why does the universe start to accelerate when the structures 
reach the non-linear regime? Or
why the present values of the 
densities of dark energy and dark matter are of the same order of magnitude? 
For a review of these and another theoretical problems on the $\Lambda CDM$ 
model, see \cite{peebles}, \cite{leandros} and \cite{bullock}.

On the other hand, there are many theoretical models that try to explain the 
accelerating expansion of the Universe, for example, among the most popular we have: 
the scalar field models (quintessence) \cite{tuner}, k-essence models \cite{almendariz}, models including dissipation\cite{winfried}, 
models inspired by the renormalization group \cite{renormalization}, modified gravity models and the 
Horndeski's model that have the  most  general  scalar-tensor action  yielding  
up to second-order equations of  motion (\cite{Horndeski}), and so forth. 
For a review see \cite{weinbergd}.

In addition, another way to understand the accelerated expansion is the phenomenological 
approach, which consists in assuming a specific functional form for the equation of state 
and studying its implications. 
The prototype of this approach is the parameterization $CPL$ \cite{cp2001,linder}. 
Other examples are: the $q(z)$ parametrization \cite{waga},
the Wetterich parameterizations \cite{wetterich}, among others e.g. \cite{parametrization}.

However, all the mentioned models to explain the accelerated expansion are dependent on the 
theory of gravitation. In this regard there is an approach that avoids it by assuming only the metric structure, 
without considering the theory of gravitation and is called the model-independent approach. 
We assumed this approach in our investigation. It is also common to call this approach as the 
cosmographic approach and derive it from a Taylor series of the scale factor 
or, equivalently, of the Hubble parameter e.g. see references \cite{weinberg2005,chiba1998,visser2003,visser2004,ck2004}.

It is worth noting that the Taylor series expansion implies fundamental difficulties with the 
convergence and the truncation of the series. Thus we can not use observational data for redshift 
greater than one, $z> 1$. For a detailed discussion of this question, see the reference \cite{visser2007}.
Consequently, in the present paper, to avoid this problem we calculate the cosmographic 
functions instead of the cosmographic parameters derived from the Taylor series.
The cosmographic functions are assumed as kinematic definitions on the scale 
factor or, equivalently, the Hubble parameter. 
In this case the cosmographic parameters correspond to the
cosmographic functions with $z=0$, i.e,  currently evaluated.

In addition, a statistical method complementary to the model-independent approach
is the method of Gaussian Processes($GP$), 
which is a non-parametric statistical method.
This means that the reconstruction of the Hubble function and its derivatives 
are performed without assuming a specific theoretical model.
Also as observational data we used the Hubble parameter data,
since from the theoretical point of view the cosmographic functions 
are directly dependent on the Hubble parameter and its derivatives. 
On the other hand, a reconstruction that uses, for example, measurements of the luminosity distance of supernovae Ia 
involves the computation of the derivative of the supernovae Ia distance data to determine the Hubble parameter and its derivatives. 
Thus this process adds a source of error propagation. This is the main reason because we only use 
Hubble parameter data in our research.

Our main objective is determine observational constraints in model-independent form using a non-parametric statistical technique, such as Gaussian processes, and observational data from the Hubble parameter. In this way we avoid using the Taylor series expansion.
We also discuss how to use these results to derive observational constraints on specific models by constructing a simple correspondence between the parameters of a given model and the cosmographic parameters. We discuss the implications of this methodology.

The present paper is organized as follows. We present the basic definitions 
in the section 2. The observational data and the simulated data
are discussed in the section 3. We summarize the method of Gaussian Processes in section 4.
We present the results of the reconstruction of our cosmographic functions in section 5
and in section 6 we give our conclusions.

\section{Cosmographic Functions}
We assume that our universe is homogenous and isotropic and therefore can be described by the 
Friedmann-Lema\^{i}tre-Robertson-Walker (FLRW) metric,
\begin{eqnarray}
 ds^{2} = -c^{2}dt^{2} +a(t)^{2}\left[\frac{dr^{2}}{1-kr^{2}}+ r^{2}\left(d\theta^{2} + 
 \sin^{2} \theta d\phi^{2}\right)\right],
\end{eqnarray}
where $a(t)$ is the scale factor and the $k$ is a constant that can have the value $= +1,0,-1$ and represent the three-dimensional curvature. 
Using this metric it is possible to introduce the cosmographic functions defined directly from the scale factor,

\begin{eqnarray}
 H(t) &=& \frac{1}{a} \frac{da(t)}{dt}, \nonumber \\  
 q(t) &=& -\frac{1}{a} \frac{d^{2}a(t)}{dt^{2}}\frac{1}{H^{2}}, \nonumber \\  
 j(t) &=& \frac{1}{a}\frac{d^{3}a(t)}{dt^{3}}\frac{1}{H^{3}},     \\  
 s(t) &=& \frac{1}{a}\frac{d^{4}a(t)}{dt^{4}}\frac{1}{H^{4}},   \nonumber \\   
 ...  &=&...                                               \nonumber \\                                          
 m_{n} &=&  \frac{1}{a}\frac{d^{n}a(t)}{dt^{n}}\frac{1}{H^{n}}.  \nonumber
\end{eqnarray}
If we evaluate these functions at the current time ($z=0$), then they are 
called the cosmographic parameters and are usually referred as
the Hubble constant ($H_{0}$), the deceleration parameter ($q_{0}$), the jerk parameter ($j_{0}$), the snap parameter ($s_{0}$), 
etc.\footnote{Curiously in the literature there are proposals of names 
for the fifth and sixth derivatives as crackle and pop (\cite{MG}).}

In the literature the traditional way to introduce cosmographic parameters is by using the Taylor expansion for the Hubble parameter
see, for example, \cite{weinberg2005} and \cite{visser2003}. However, the use of observational data and the Taylor series leads to two main inconsistencies. The first arises when we use observational data with $z>1$, since from the mathematical point of view the series has a convergence radius at the most $ |z| = 1 $, thus it is only justified to use data with redshift $ z << 1 $.
Otherwise the errors would increase and the series would not be useful to represent the original function. 
The other inconsistency arises when a new redshift variable is introduced to transform the domain of redshift values and to be able to use data with $ z> 1 $, the problem is that there is no one-way to do this. Furthermore, a given redshift parametrization contributes to error propagation. In what follows we briefly review these two inconsistencies.

\begin{itemize}
\item Taylor series and their error in cosmological models: the Hubble's function can be expressed in Taylor series as a function of the redshift using the relation 
$1+z=\frac{a_{0}}{a(t)}$ \cite{weinberg2005},

\begin{eqnarray}
 H(z) = H_{0} + \frac{dH}{dz}\Bigg|_{0} z + \frac{1}{2!}\frac{d^{2}H}{dz^{2}}\Bigg|_{0} z^{2} + 
 \frac{1}{3!}\frac{d^{3}H}{dz^{3}}\Bigg|_{0} z^{3}  + O(z^{4}).
\end{eqnarray}

The series is written around the current redshift, $z = 0$, and mathematically represents the approximate 
Hubble's law for an arbitrary order.
\footnote{On the generality of the equation above it is important to mention that for gravitation theories of high derivative 
one can not reconstruct his equation of state without making use of an external hypothesis, because these theories include new 
degrees of freedom. For details see references \cite{dombriz2016} and \cite{Busti2015}.}
In this context, we consider the truncation error. 
To briefly illustrate this question we have used three cosmological models with 
Hubble's law exact, so that we can easily determine its Taylor series.
We consider the flat $\Lambda CDM$ model, the $CPL$ model and a $q(z)$ parametrization.
In all case we consider the flat universe. \\
Thus, for the case of the flat cosmological constant model the Hubble parameter can be written,

\begin{eqnarray}
 H(z) &=& H_{0} \left[\Omega_{m0}(1+z)^{3} + (1-\Omega_{m0}) \right]^{1/2},
\end{eqnarray}
where $\Omega_{m0}$ is the the matter density parameter.\\ 
On the other hand, for the CPL (Chevallier-Polarski-Linder) parametrization \cite{cp2001} and \cite{linder} we have,

\begin{eqnarray}
 H(z) = H_{0}[\Omega_{m0}(1+z)^{3} +(1-\Omega_{m0})(1+z)^{3(1+w_{0}+w_{1})}
e^{3w_{1}(\frac{1}{1+z}-1)}]^{1/2}, 
\end{eqnarray}
where $w_{0}$ and $w_{1}$ are the parameters of the model and although it 
does not behave very well for high redshift it allows reproduce the Hubble law and the distances of different 
models of dark energy with very good precision and in the literature it has been intensively used, 
see for example \cite{padmanabhan} and \cite{scherrer} and references therein.
This model represents dynamic dark energy.\\

We are also consider another model of dynamic dark energy with four free parameters \cite{waga} for which have 

\begin{eqnarray}
 H(z) = H_{0}\left[(1+z)^{(1+q_{i})}\left(\frac{q_{i} 
 \left(\frac{1+z_{t}}{1+z}\right)^{1/\tau}-q_{i}}{q_{i}(1+z_{t})^{2(1+q_{i})}-q_{i}}\right)^{\tau(q_{i}-q_{f})}\right],
\end{eqnarray}
where $q_{i}$ and $q_{f}$ are the initial and final deceleration parameter respectively and $z_{t}$ is the 
transition redshift from a decelerated universe to an accelerated universe and $\tau$ is the width of the transition.
This model is quite generic and allows to study in greater detail the transition for an accelerated universe. 
It is also interesting to know that this parametrization includes several other models of dark energy, for details and observational constraints of the parameters see \cite{vargas}.

\begin{figure*}
	\includegraphics[width=3.4in,height=3.0in]{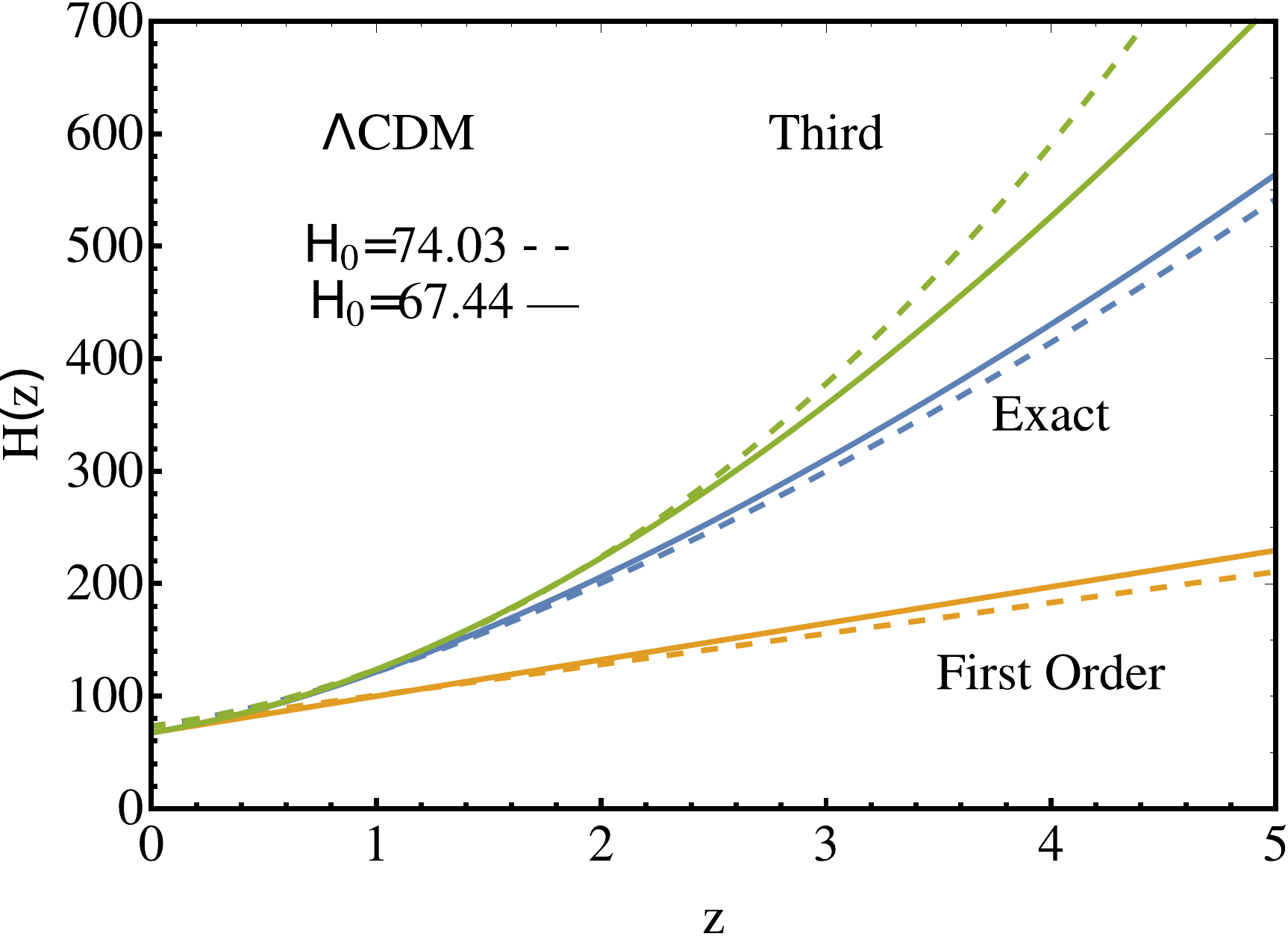}
	\includegraphics[width=3.4in,height=3.0in]{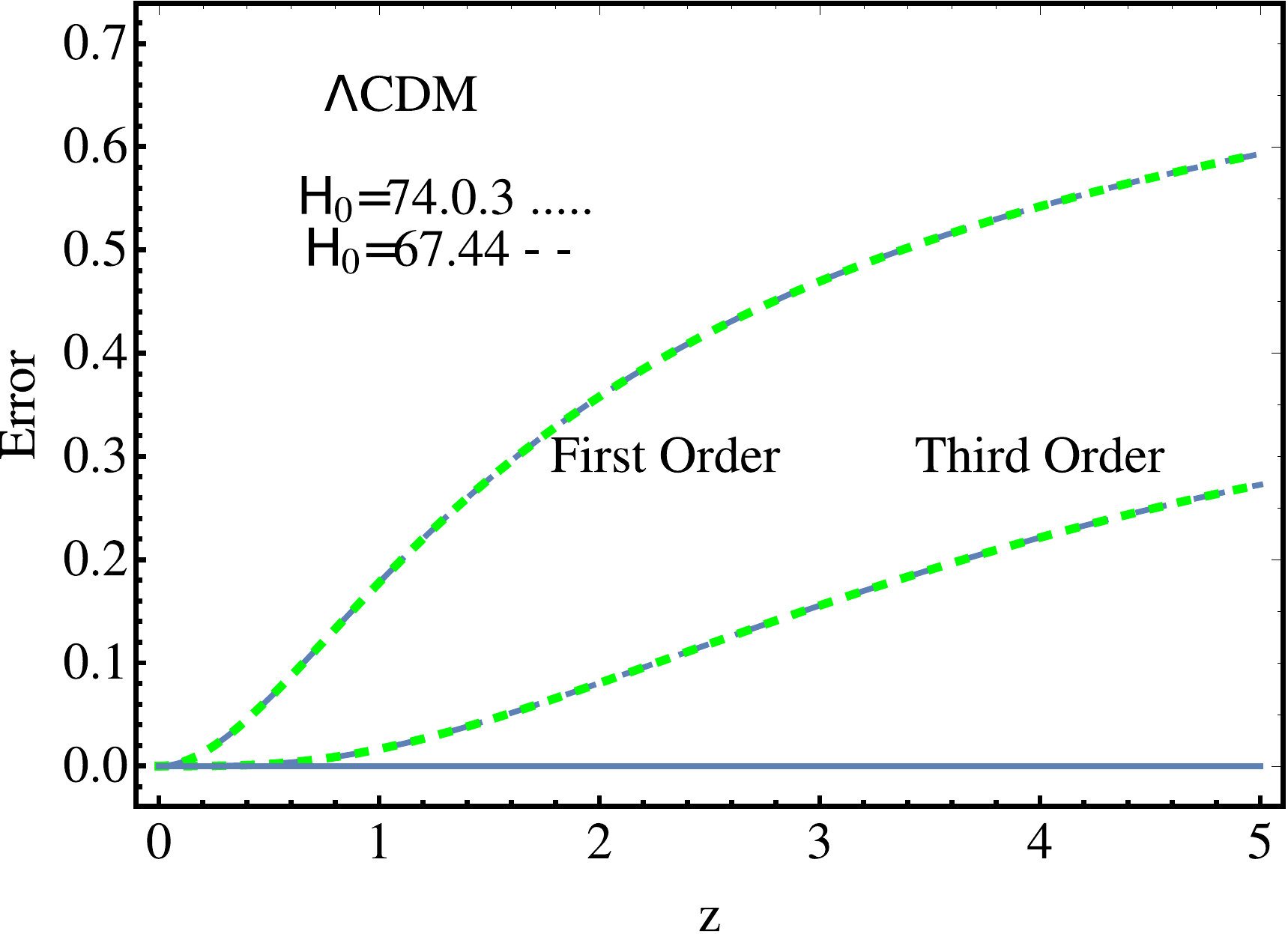}
	\includegraphics[width=3.4in,height=3.0in]{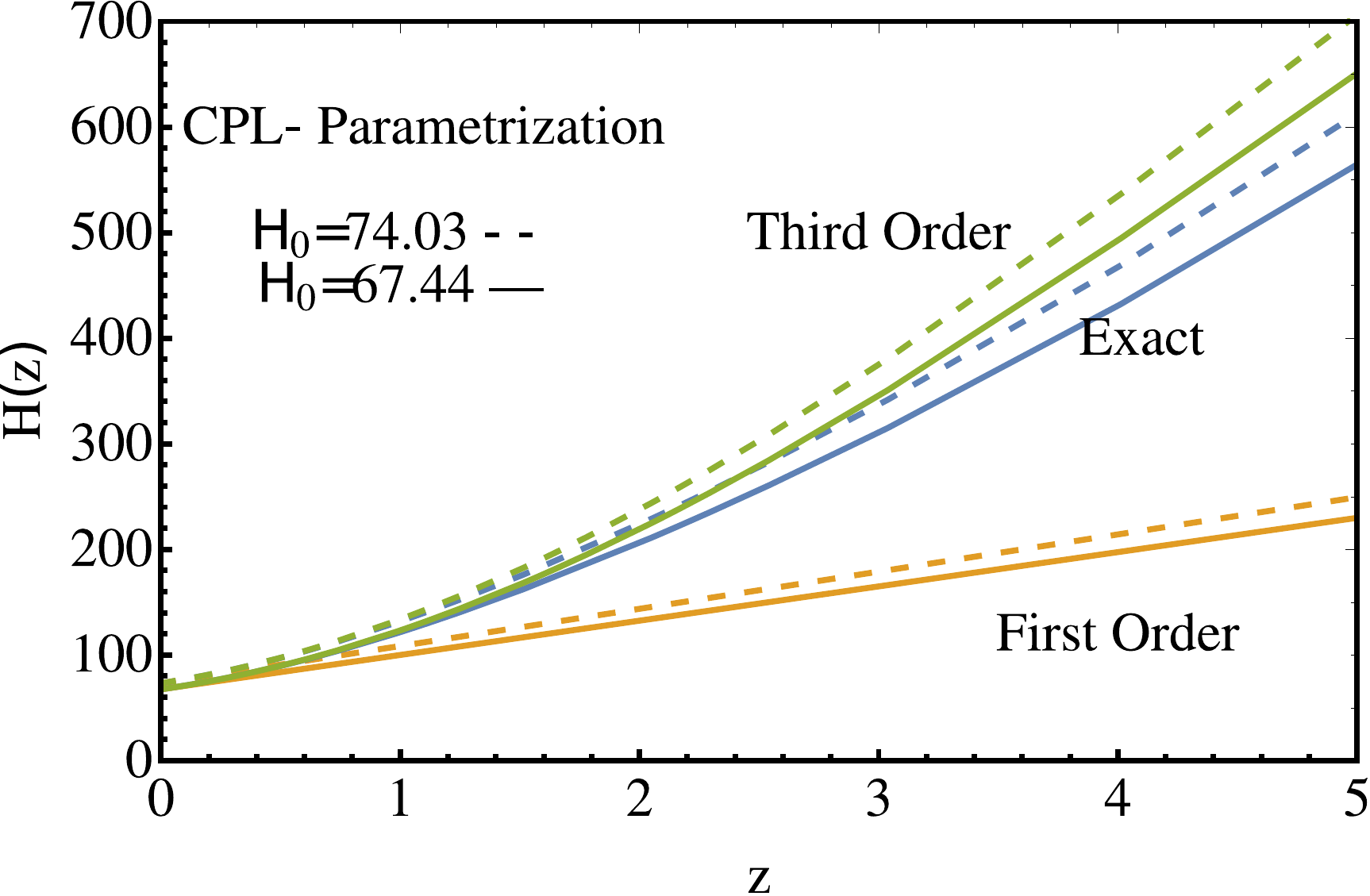}
	\includegraphics[width=3.4in,height=3.0in]{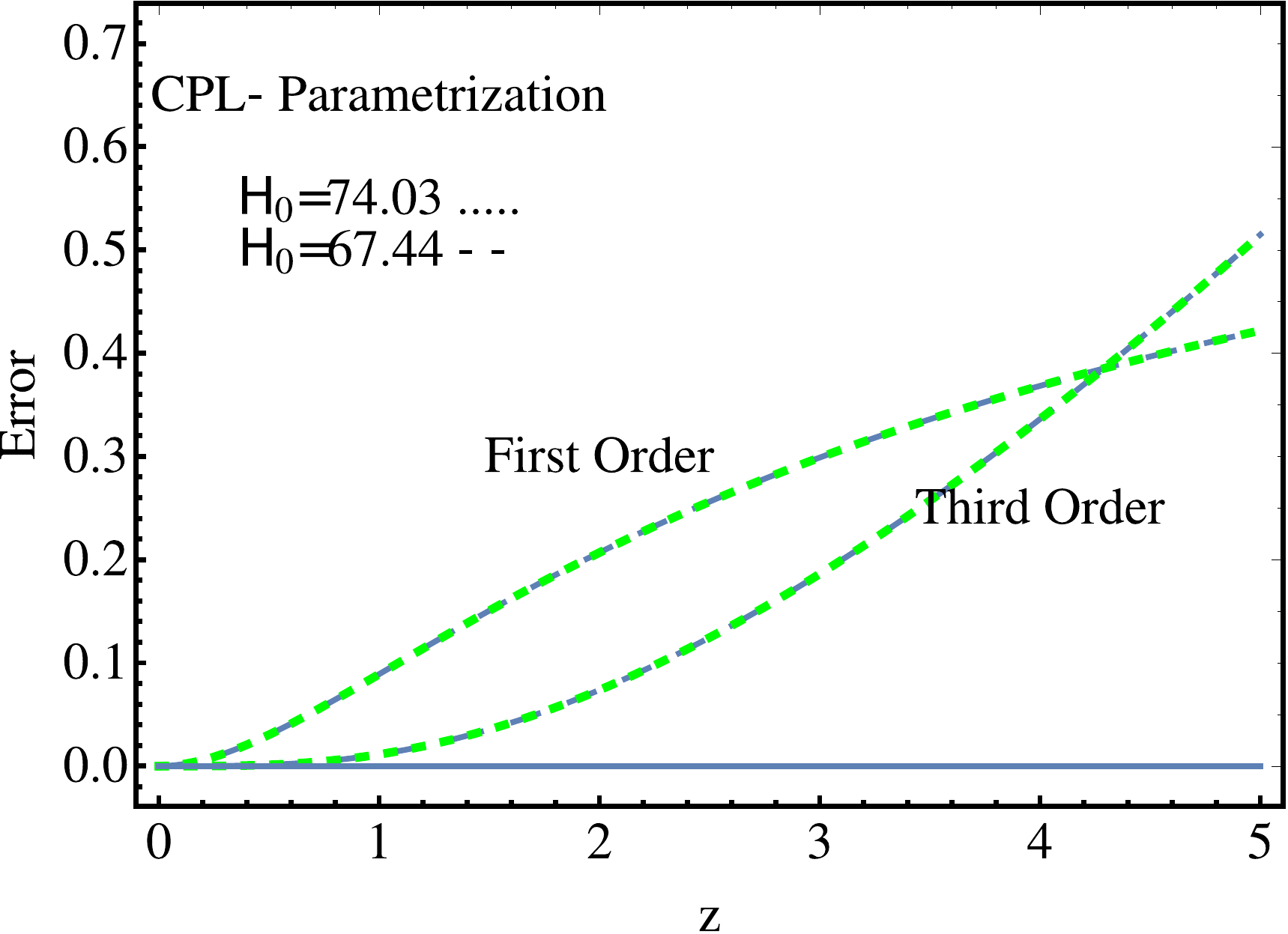}
	\includegraphics[width=3.4in,height=3.0in]{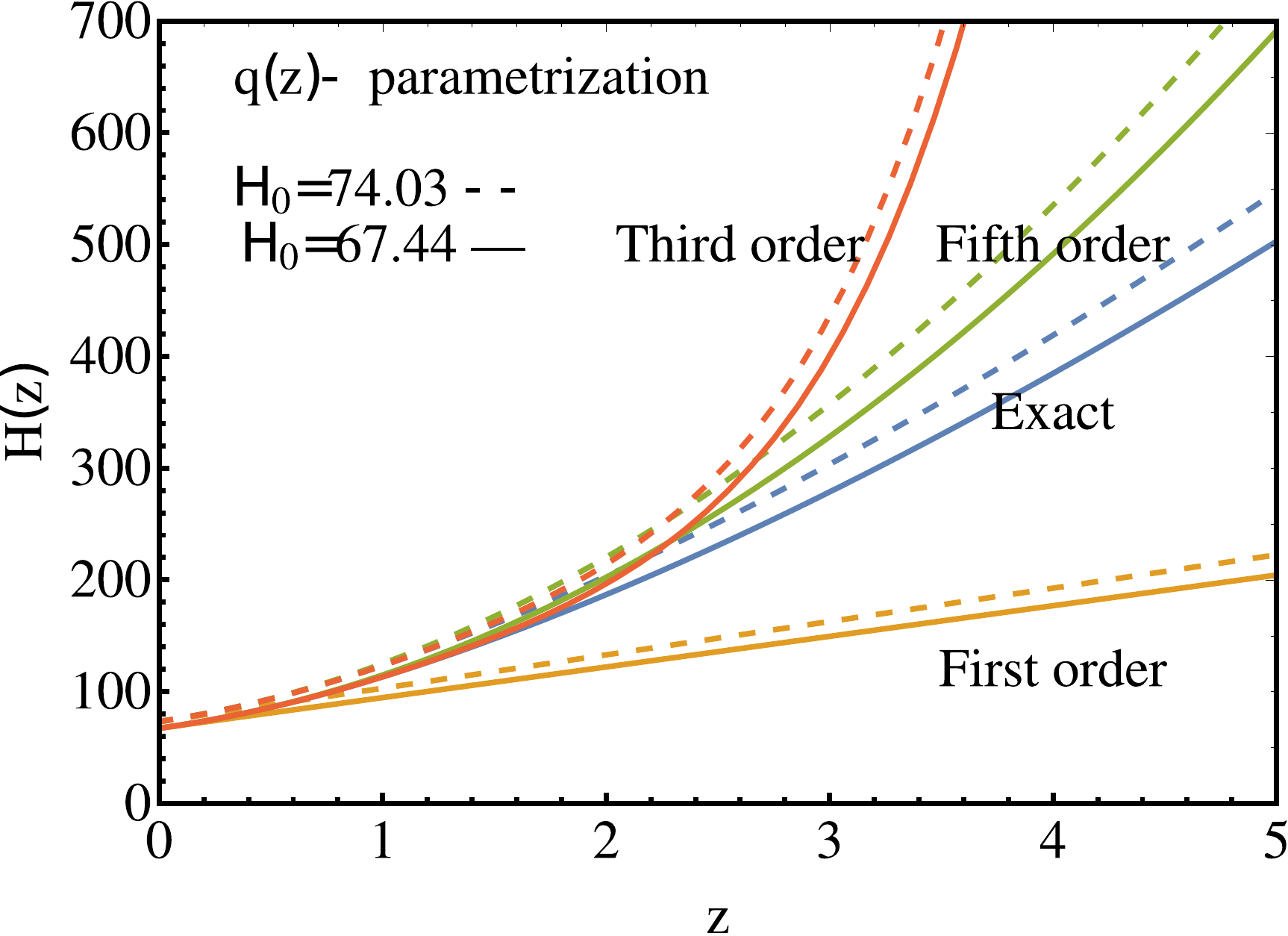}
	\includegraphics[width=3.4in,height=3.0in]{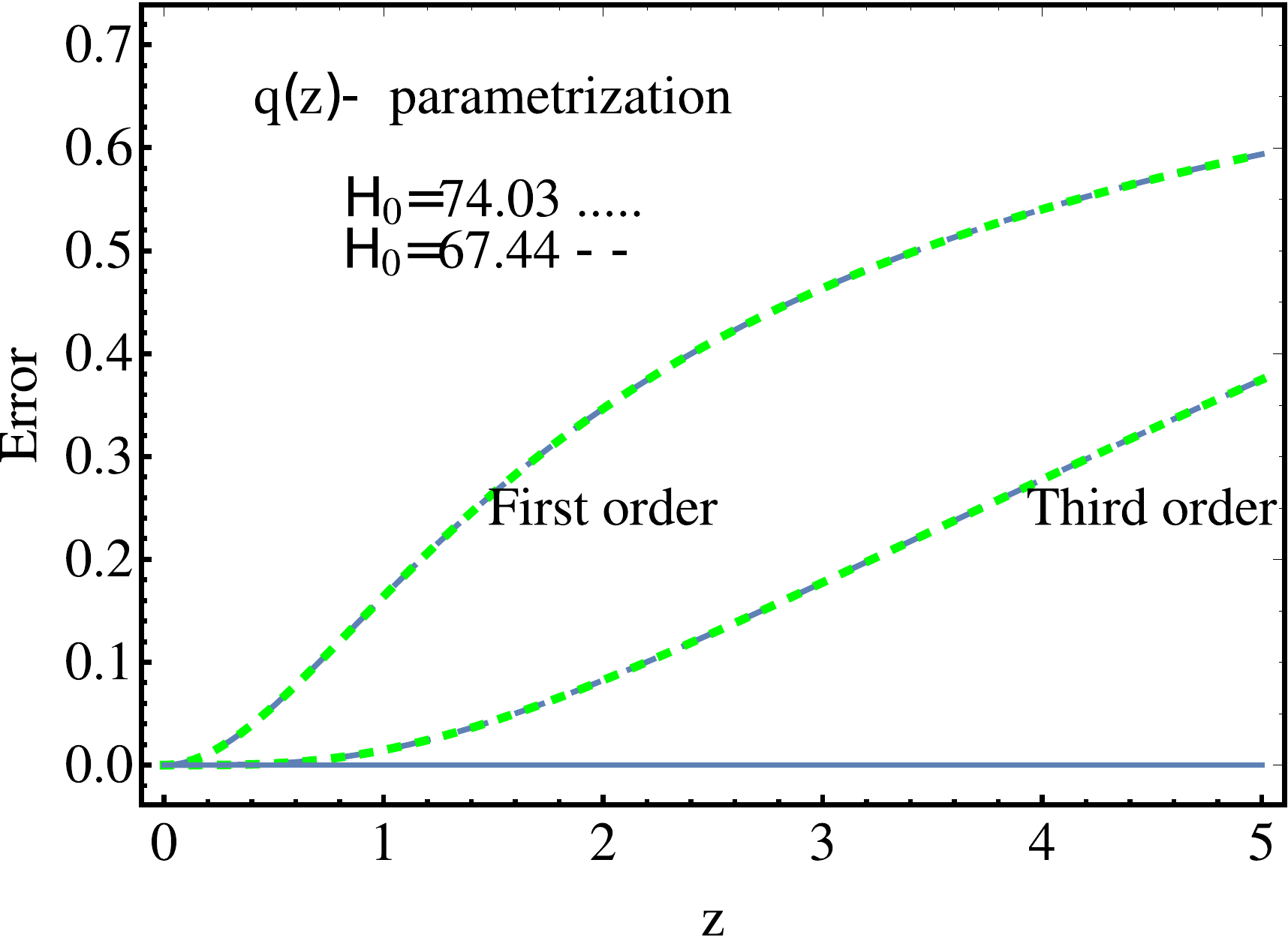}
    \caption{In the top left the Hubble function for $\Lambda CDM$, in the middle left for the CPL model and in the 
		botton left for the $q(z)$ model and in the right side the quantification of the error respectively for each model. 
		In all case we use $\Omega_{m0}=0.32$ and we consider two values for the today Hubble parameter:
		the highest value corresponds to the SH0ES Collaboration and the lowest value corresponds to the Planck Collaboration.}
\end{figure*}

All these models have an analytical expression for the Hubble parameter and have been constrained observationally. 
We calculate the Taylor series for each of these models up to the third power in the redshift variable. 

In Figure 1, we show the exact Hubble function versus the third order approximation.
We can observe that up around $z=0.5$ the values between the exact function and the third order approximation are more or less equivalent.
To quantify explicitly this difference we plot the magnitude of the error defined as:

\begin{eqnarray}
Error &=& \frac{H(z)_{exact}-H(z)_{approximate}}{H(z)_{exact}}.
\end{eqnarray}

Also in figure 1, right column, we can see 
that the error between the exact function and the third-order 
function for the range of redshift, $ 0 <z <1 $, is relatively small, but for higher values of the redshift the error increases considerably.
Therefore, the use of the Taylor series to compute observational constraints for data with $ z> 1 $ leads to considerable truncation errors. One way to circumvent this problem could be to select observational data with $ z < 1 $, this would give consistent results within the truncation error of the Taylor series.

\item New redshift variable:
a widely used proposal in the literature on the truncation problem is to use a new variable for the redshift. The initial and best motivated proposal is given by the reference \cite{visser2007} and establishes the following relationship
\begin{eqnarray}
y&=&\frac{z}{1+z},
\end{eqnarray}
where $ y $ is the new redshift variable. This new redshift variable allows mapping the redshift values with $ z> 1 $ to a region with $ y <1 $ and can be defined as the change of the emission wavelength divided by the observed wavelength,
\begin{eqnarray}
1-y&=&\frac{\lambda_{e}}{\lambda_{0}} = \frac{1}{z}.
\end{eqnarray}

The correspondence between the values of redshift $z$ and the values of the variable $y$ are: in the past $ z \in (0, \infty) $ and corresponding the values $ y \in (0,1) $; in the future $ z \in (-1,0) $ and corresponding to the values $ y (\infty, 0) $. 

However, this redshift variable transform is not unique 
in the literature there are other proposals such as \cite{aviles2012},
\begin{eqnarray}
y_{1} &=& \arctan(z),\\
y_{2} &=& \arctan(\frac{z}{z+1}),\\
y_{3} &=& \frac{z}{1+z^{2}}. 
\end{eqnarray}

In this sense, the cosmological constraints depend on the parametrization assumed for the redshift.
When comparing the results using the variable $y$ with those obtained using the original redshift variable inconsistencies arise. 
It is quite instructive to show this explicitly. For example, we can consider the expressions for
the luminosity distances in the variables $ y $ and $ z $ respectively \cite{weinberg2005} and \cite{visser2007},
\begin{eqnarray}
d_{L}(y) &=& \frac{c}{H_{0}} y [1-\frac{1}{2}(-3+q_{0})y+ \frac{1}{6} (12-5q_{0}
 +3q_{0}^{2}+6(j_{0}+ \Omega_{0}))y^{2} + O(y^{3})], \\
d_{L}(z) &=& \frac{c}{H_{0}} z[1+\frac{1}{2}(1-q_{0})z+ \frac{1}{6}(1-q_{0}
 -3q_{0}^{2}+j_{0} )z^{2} + O(z^{3})].
\end{eqnarray}

Now we can use these two expressions to determine observational constraints on the parameter space $( q_{0} $, $ j_{0} )$. 
To determine these constraints we can use a robust data set from the point of view of error control, 
particularly systematic errors. We use data from type Ia Supernovae, specifically the 
Pantheon sample with 1048 supernovae \cite{scolnic}. 
In appendix A we briefly describe the method used to estimate parameters using these data. 

In figure 2 we show the results of the confidence contours for the parameters space $(q_{0},j_{0})$
using the type Ia Supernovae data.
On the left side we present the results for the original redshift, $z$, with the best fit 
$q_{0}=-0.579 \pm 0.103$ and $j_{0}=0.977 \pm 0.232$ with $1 \sigma$. 
On the right side we show the results for the redshift, $y$, 
with the best fit $q_{0}=-1.107 \pm 0.103$ and $j_{0}=0.650 \pm 0.232$ with  $1 \sigma$.
As we can see, the results are incompatible since that the confidence contours do not overlap.\footnote{Similar results are obtained
in a recent paper by \cite{capozziello} also using type Ia Supernovae.}

\end{itemize}

Therefore, to avoid the problems mentioned about the Taylor series we propose to use directly 
the definitions of cosmographic functions as given above. Thus, for our analysis it is convenient to rewrite 
the cosmographic functions explicitly as a function of the Hubble parameter and its derivatives with respect to the redshift,
\begin{align}
 q(z) &=  (1+z)\frac{H'}{H} - 1 \\
 j(z) &= \frac{H''}{H} (1+z)^{2} + \left(\frac{H'}{H}\right)^{2} (1+z)^{2} -2 \frac{H'}{H}(1+z) + 1 \\
 s(z) &= -\frac{1}{2}(1+z)^{2} \frac{[H^{2}(z)]'''}{H^{2}(z)} -\frac{1}{2} (1+z)^{2} \frac{[H^{2}(z)]'' H'(z)}{H^{3}(z)} \\&
+ \frac{(1+z)^{2} [H^{2}(z)]''}{H^{2}} + \frac{(1+z)^{2} [H^{2}(z)]' H'(z)}{H^{3}(z)} 
+ \frac{3(1+z)H'(z)}{H(z)} + 1, \nonumber
\end{align}
where the prime represents derivative with respect to redshift. Furthermore, using the snap parameter definition we have 
$ \ddddot{a} = s a H^{4} $, from the jerk parameter we have $\dddot {a} = j a H^{3}$ and of the deceleration parameter we have $\ddot{q}=-q aH^{2}$, thus we can write a useful relationship between these cosmographic functions given by the expression \cite{jerk},
\begin{eqnarray}
s(z) &=& -(1+z) \frac{d j(z)}{dz} - j(z)\left(2+3q(z)\right).
\end{eqnarray}
It is also useful to write the cosmographic parameters for the flat $ \Lambda CDM $ model,
\begin{eqnarray}
q_{0} &=& -1 + \frac{3}{2} \Omega_{m0} \\
j_{0} &=& 1 \\
s_{0} &=& 1- \frac{9}{2} \Omega_{m0}.
\end{eqnarray}
As we can see for the flat $ \Lambda CDM $ model  these cosmographic parameters are 
defined by the value of the matter parameter today. These results are 
useful to make a comparison with our results.
We use the set of equations (16) - (18) to reconstruct the cosmographic functions, i.e, $q(z)$, $j(z)$ and $s(z)$ 
using observational data from the Hubble parameter and simulated data.

\begin{figure*}
        \includegraphics[width=3.00in,height=3.00in]{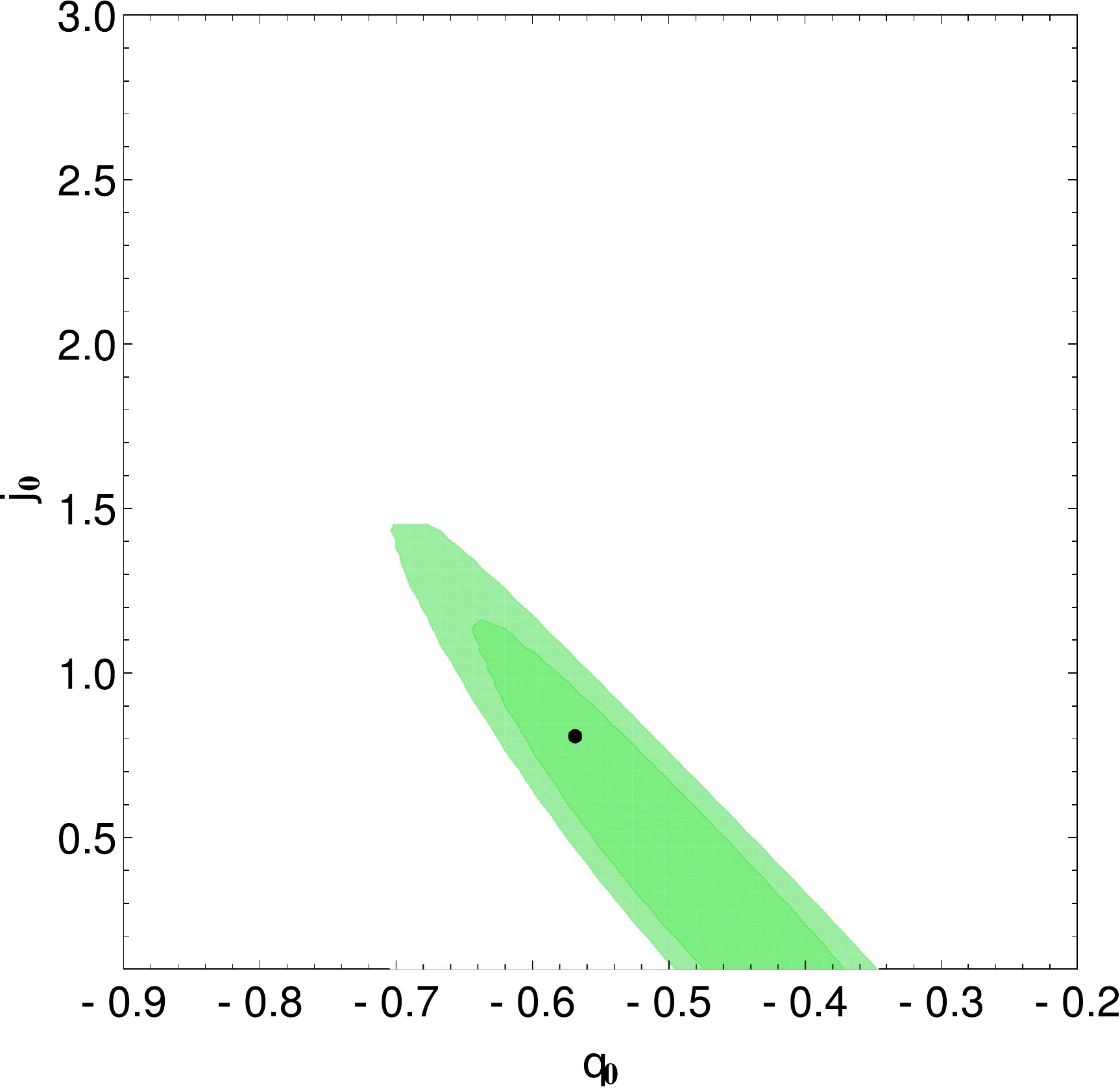}
	\includegraphics[width=3.00in,height=3.00in]{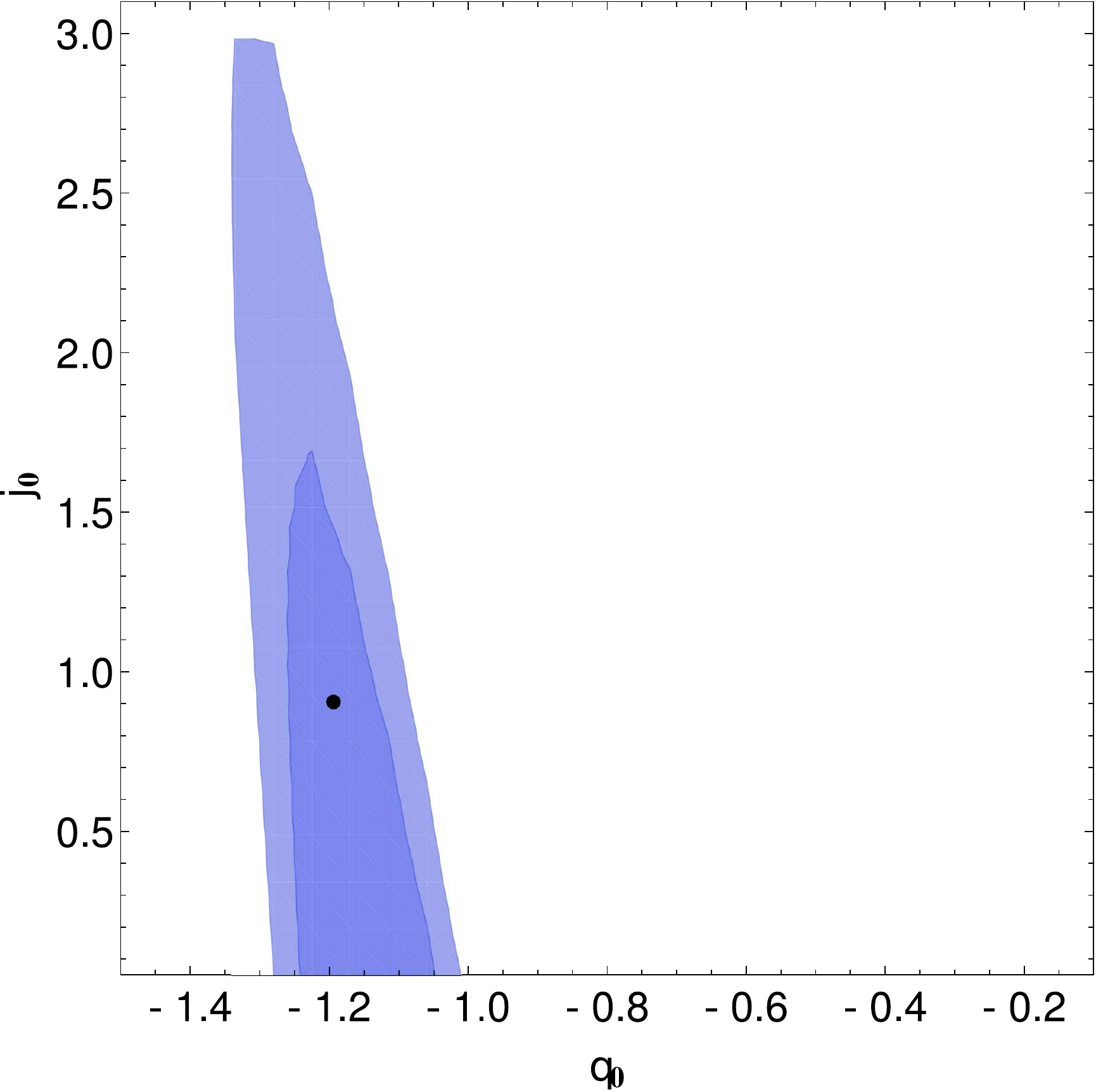}
  \caption{In the left we shown the observational constraints on the parameters $q_{0}$ and $j_{o}$ using the redshift (z) and in the right shows the observational constraints using the $y$-redshift. We can see that both results are incompatible. 
	The data used to determine these observational constraints have been data from Type Ia supernovae, for details see appendix A.}
\end{figure*}

\section{Data}
As we have observed in the previous section the cosmographic functions can be written 
directly as a function of the Hubble parameter and its derivatives. 
This allows us to discern that the most appropriate data to minimize error propagation are the Hubble data.
Therefore, we focus on this class of data. 
\subsection{Observational Data}
There are two efficient and widely used forms to obtain Hubble parameter measurements.
\subsubsection{Cosmic Chronometers(CC)}
This method is based on the expression of the differential age of the
universe as a function of redshift,
\begin{eqnarray}
 H(z) &=& -\frac{1}{1+z}\frac{dz}{dt}.
\end{eqnarray}
This method was proposed by \cite{loeb} and consists in directly measuring the amount $ dz/dt $ and, consequently, 
the Hubble parameter. The most used data to measure this amount have been passively evolving galaxies 
with high-resolution spectroscopic data along with synthetic catalogs to limit the age 
of the oldest stars in the galaxy. A complete description of this methodology can be reviewed 
for the $SDSS$ and $GDS$ data in the reference \cite{jimenez}.

\subsubsection{The Radial BAO Size Method}
The method is based on measurements of the acoustic scale of $BAO$,
which is more accurate with respect at the $CC$ method \cite{gaztanaga}.  
This accuracy is understandable because $BAO$ mainly depends on a spatial  
measurement compared to the first method where a time measurement is  required which 
increases systematic errors. However, this method of $BAO$  requires 
assuming a prior in the radius of the sound horizon. Hence, we have

\begin{equation}
 H(z) = -\frac{r_{bao}}{r_{cmb}}H_{fiducial}.
\end{equation}

This method depends on the fiducial model which is assumed to be the flat $\Lambda CDM$ model.

On the other hand, in the literature there are different compilations of samples of the Hubble parameters data. We use the sample presented by \cite{Zheng} what includes data of both $CC$ and $BAO$, but we not including the points $z = 0.4497$ and $z = 2.34$. The first point is excluded because it overlaps with another data and has a negligible effect on the results. The other point has a strong influence on the estimation of parameters and because it is very restrictive we have excluded it. Therefore our sample has 34 points. This sample has been used in \cite{velasqueztt} showing results compatible with the literature.

In figure 3 we show the data used. We can see that the errors of the $ CC $ data can reach up to a value of
$ \Delta H \approx 62 Km \: s^{-1}  Mpc^{-1} $ and the errors of the BAO data have a mean of about $ 7 Km \: s^{-1}  Mpc^{-1} $.

Before using these data it is important to verify their Gaussian nature \cite{podariu}. 
To verify this we determine the factor $ N $ which must be small to be consistent with the gaussianity hypothesis. 
Briefly the methodology used is shown in Appendix B. The results are presented in Table 1 and we can see that the 
data are compatible with the gaussianity hypothesis. Therefore, we have a suitable sample to use the Gaussian process method.

\subsection{Simulation Data}
Our simulation follows the prescriptions established in \cite{mazh}:
we choose as background cosmology the flat $ \Lambda CDM $ model using the values 
$ \Omega_{m0} = 0.315 \pm 0.001$ of the \cite{planck} and $ H_{0} =69.45 \pm 4.34 $( best fit of $H (z)$ observational data).
Then, with these values we determine the $H_{fiducial}$ using the equation (5). We assume that the deviation of the simulated value 
with respect to fiducial value is calculated as: $\Delta H = H_{sim} - H_{fiducial}$, where the $\Delta H$ can be derived from a Gaussian distribution of the form $ N (0, \tilde{\sigma}) $, where the $ \tilde {\sigma} $ is a random number 
which also can be drawn from the following Gaussian distribution $N\left(\sigma_{0},\varepsilon(z)\right)$.
In this Gaussian distribution we introduce the information about the error as follows: 
we define the $ \varepsilon(z) $ as $ \varepsilon(z) = \frac {\sigma_{+}(z) - \sigma_{-}(z)} {4} $ 
and the parameter $ \varepsilon(z) $ is chosen so that the probability of $ \tilde{\sigma} $ falling within 
the strip of $ 95.4 \% $. We also define the $ \sigma_{0} = (\sigma_{+} + \sigma_{-}) / 2 $.
We assume a linear function for the $ \sigma $ that allows us to model the simulated errors. This can be deduced from figure 3 where we can see that, except for some special cases, the error increases with the redshift. 
Thus, in general we define $\sigma_{\pm} = a_{\pm}z+b_{\pm}$ where $a$ and $b$ are chosen according to the errors of the Hubble data
and the $+$ sign indicates the upper line and the $-$ sign the lower line as shown in the lower part of figure 3. 
For example, for the case of $CC$ we assume the straight lines (orange in the figura 3):  $ \sigma_{+} = 19.999z + 11.105$  and $ \sigma_{-} = 3.650 z + 1.65$ and in the case of $BAO$ the blue lines.

\begin{table}
\caption {Gaussian nature of the data} \label{tab:title} 
\centering
\begin{tabular}{ccccc}
\\ 
\hline
\hline
$N$ & $\bar{z}$ & $\bar{H}$ & $\bar{\sigma}$ & $N_{\sigma}$  \\ 
\hline
\hline
2 & $0.084$  & $69$ & $10.2342$ & $1.41$ \\
5  & $0.184$  & $75.939$ & $2.878$ & $1.443$ \\
4  & $0.332$  & $81.785$  & $6.309$	& $1.861$ \\
5  & $0.417$  & $83.232$  & $4.824$	& $1.429$ \\
4  & $0.516$  & $88.687$  & $5.341$	& $0.926$ \\
3  & $0.624$  & $91.191$  & $4.544$	& $0.624$ \\
3  & $0.759$ & $102.146$  & $5.696$	& $0.164$ \\
3  & $0.965$  & $132.079$ & $14.120$	& $0.34$ \\
5  & $1.445$  & $160.331$ & $8.717$	& $0.088$ \\
2  & $2.333$  & $221.328$ & $6.933$	& $0.427$ 
\\ 
\hline \hline
\end{tabular}

\end{table}

\begin{figure*}
        \includegraphics[width=3.900in,height=3.00in]{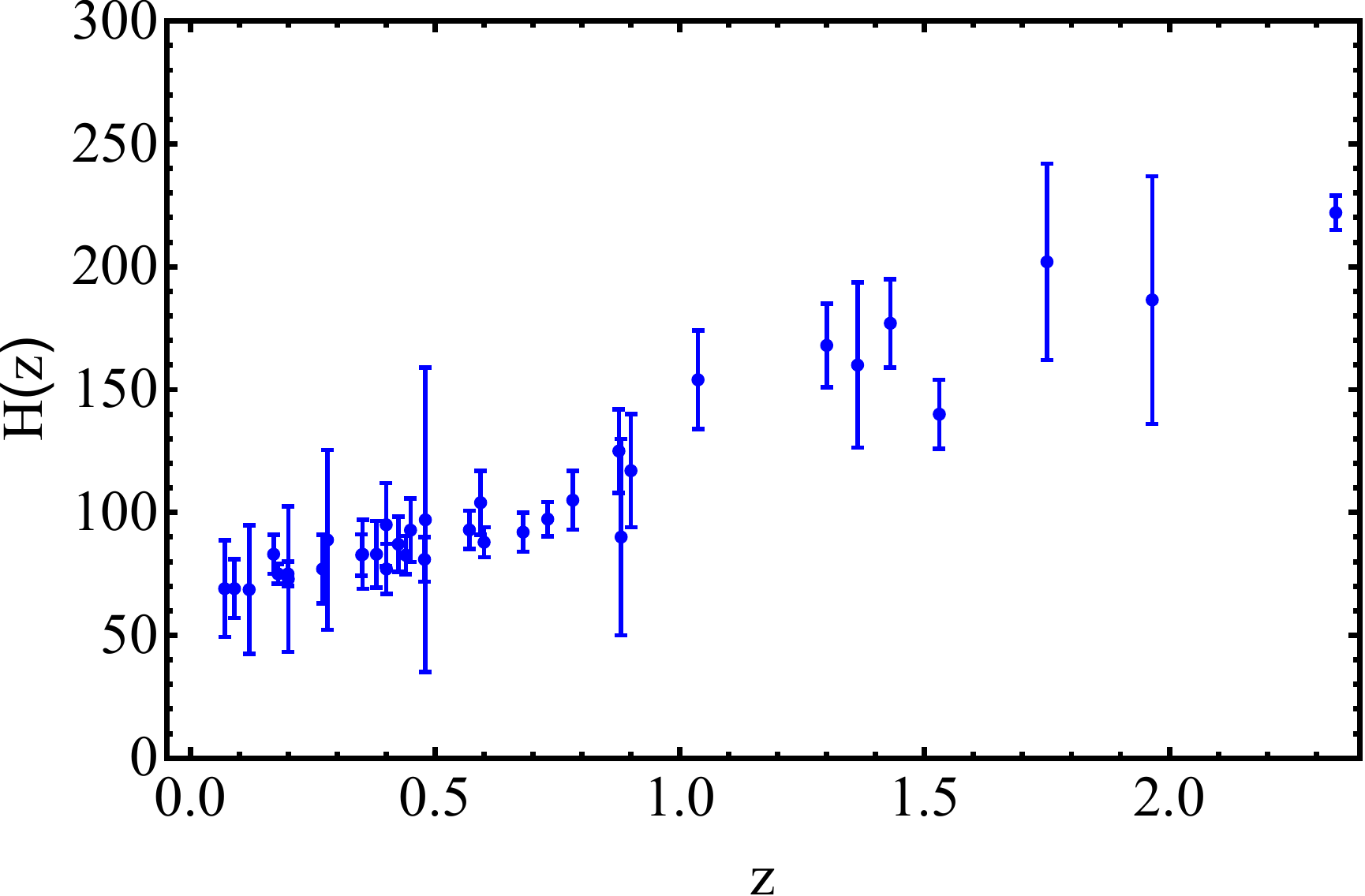}
	\includegraphics[width=4.00in,height=3.00in]{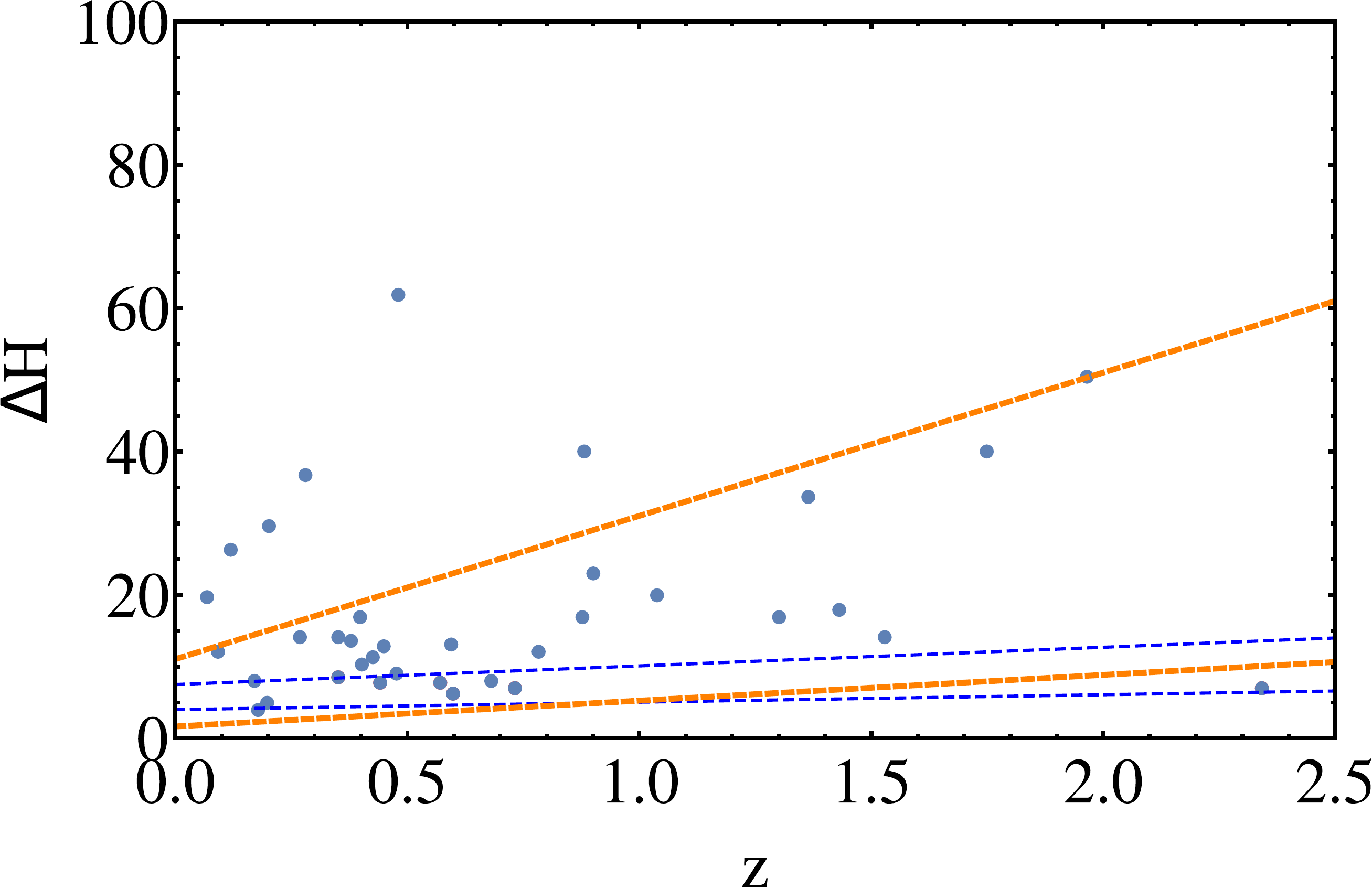}
    \caption{We shown the Hubble data set (CC+BAO) used in this investigation and also the lines used to calculate the error of simulated data.
		The orange lines represent the lines for the CC data and the blue lines represent the lines for the BAO data.}
    \label{fig:example}
\end{figure*}

\section{Gaussian Processes}

The gaussian processes ($GP$) are a nonparametric statistical method 
allowing the reconstruction of a given function directly from the manipulation of data. 
They can be thought as a extension to a multivariate normal distribution, 
that is, a $GP$ is a infinity collection of random variables, but where every 
finite set have a normal distribution. A $GP$ is a distribution over 
functions, but in the practice a $GP$ allows us to derive a posterior 
distributions by simply considering a finite set of points that are associated with the observational data.
For details see \cite{rasmussen}. A $GP$ can be written as:
\begin{eqnarray}
f(x) \sim GP\left(\mu(x),k(x,\tilde{x})\right),
\end{eqnarray}
where the value of $f$ when evaluated at a point $x$ is a gaussian random variable with
mean $\mu(x)$. Additionally, the value of the function $f$ at the point $x$ is not independent of the value of the 
function $f$ at some other point nearby $\tilde{x}$, but is related 
by the covariance function $k(x,\tilde{x})$. 
Considering observational data $(x_{i},y_{i})$, assuming that the errors are gaussian
and that $y_{i}=f(x_{i}) \pm \sigma_{i}$, where $\sigma_{i}$ are the $1\sigma$ error bars and $i=1,...,N$, we can to
reconstruct the function $f$ at chosen points.
This function is denoted by $f^{*}$. This reconstruction can be done through 
the joint distribution between $f^{*}$ and $y_{i}$. For details of the exact expressions see the reference \cite{seikel2012}.
In general, this reconstructed function has a reconstructed mean given by,
\begin{eqnarray}
\bar{f}^{*}(x) = \sum_{i,j=1}^{N}{k(x,x_{i})(M^{-1}})_{i,j}(f(x_{j})-\mu(z))
\end{eqnarray}
in our reconstruction we chosen as prior mean function $\mu(z)$ a constant value and for the standard deviation:
\begin{eqnarray}
\sigma(x) = k(x,x)-\sum_{i,j=1}^{N}{k(x,x_{i})(M^{-1}})_{i,j}k(x_{j},x)
\end{eqnarray}
where $M_{ij}=k(x_{i},x_{j}) + c_{ij}$ and $c_{ij}$ represents the covariance of the input data. 

For our calculations, we assume the exponential
function as a covariance function which is given by
\begin{eqnarray}
k(x,\tilde{x})=\sigma _{f}^{2}\exp \left(-\frac{(x-\tilde{x})^{2}}{2\ell^{2}}\right),
\end{eqnarray}
where $\sigma_{f}$ and $\ell$ are called hyperparameters which are determined by maximizing
the log marginal likelihood, see \cite{rasmussen},
\begin{eqnarray}
\ln \mathcal{L} = -\frac{1}{2} \sum{\left[f(x_{i})-\mu(x_{i})\right](M^{-1})_{i,j}\left[f(x_{j})-\mu(x_{j})\right]}
- \frac{1}{2}\ln |M| -\frac{1}{2} N \ln 2 \pi, 
\end{eqnarray}
where $M$ represents the determinant of $M_{ij}$. It is important to note that instead of optimizing these hyperparameters, they can be marginalized by using, for example, the MCMC method. However, a simple method of testing the validity of the optimization process is to vary the initial value of the hyperparameters and see if the values obtained change significantly. We have done this process and we determine that our hyperparameters do not change significantly. Therefore, we are confident in using the optimization process. On the other hand, the $MCMC$ method should be used to obtain high precision restrictions associated with high quality data.

The question of how to choose a suitable covariance function is an important problem for GP.
In the present investigation we show some evidence to consider the Gaussian nature of the data, see table I. However, we consider that this evidence is not conclusive. Some papers as \cite{busti,ratrar} have used other covariance functions to estimate $ H_{0} $ and have shown that within $1 \sigma$ different choices are consistent.
Therefore, in the present paper we only use the squared exponential covariance function and we leave the specific study of the effects of choosing different covariance functions for further investigation.

Additionally, the GP allows to reconstruct the derivative of the data. To implement this method, we use 
the public package $GaPP$ \cite{seikel2012}. 
For applications of GP in cosmology consider references \cite{seikelb} and \cite{amvt} and for another GP methodology to see 
\cite{Holsclawprl, Holsclawpr, Holsclawpr2011}.

\section{Results}
In figure 4 we present the reconstruction of the 
Hubble parameter using the $GaPP$ code. In the top we present the model-independent reconstruction of $ H (z) $, where we have used the 
$CC + BAO$ data to determine the Hubble constant ($ H_{0} = 69.45 \pm 4.34$). In the other two figures we reconstructed $H(z)$ using as prior
for the Hubble constant the values of the Planck and SH0ES Collaborations. In all cases, the reconstruction is done
with $1\sigma$ of uncertainty. In figure 5 we can observe the reconstruction of the cosmographic functions $ q (z) $, $ j (z) $ and $ s (z) $
without using a prior value for the Hubble constant.
In particular the deceleration parameter shows that the transition redshift, $z_{tr}$, from a decelerated universe to an accelerated universe, defined as $ q(z_{tr})=0$, is into the region $z <1$ with $ 2\sigma$.

We obtain the values: $z_{tr}=0.637^{+0.165}_{-0.175}$ (model-independent) and $z_{tr}=0.670^{+0.210}_{-0.120}$ 
with $H_{0}=67.44$ (Planck) and $z_{tr}=0.710^{+0.159}_{-0.111}$ with $H_{0}=74.03$(SH0ES). 
In accordance with these results we can conclude that $ z_{tr} \in \left [0.550, 0.870 \right] $.
These values for, $z_{tr}$, are in accordance with other results determined in the literature, for example, see 
\cite{farooq2013a,farooq2013b,farooq2017,yu2018,mamon,amvt2011}.

On the other hand, for values $ z > 1 $ the function $ q (z) $ 
shows a trend to remain in the region $ q> 0 $ associated with a phase dominated by matter. 
However, for high redshift values we see a abrupt drop which can be associated with the few observational data for this region  (see figure 3).

In figure 6 we can see that the reconstructed function $ j(z) $ does not exclude the flat $\Lambda CDM $ model  with $ 2\sigma$, however the average value is different from the flat $\Lambda CDM $ model for all the reconstruction.
The figure shows a trend for $ j <0 $ when $ z> 1 $, this is due to the propagation of errors for high redshift of the derivatives of the Hubble parameter. For the case of the Hubble constant of the Planck collaboration the value of $ j_ {0} $ is close to the value of the flat $ \Lambda CDM $ model.

With respect to the snap parameter if we consider the value of the matter density parameter measured by the Planck collaboration, $ \Omega_{m0} = 0.315 $, then for the flat $ \Lambda CDM $ model implies a value of $ s_{0} = -0.4175$. Our results with $ 1 \sigma $ are marginally in agreement with this value (see values in table 2). Secondly when $ z> 1 $ the error increases remarkably because reconstructing the parameter 
$s(z)$ implies reconstructing third-order derivatives for the Hubble parameter.

In figure 7 we show the same results as figure 6 but for simulated data.
On the left side,  we use simulated data with the error prescription from the $CC$ technique and on the right side we use the simulated data compatible with the error from the $BAO$ technique. In both cases we observe that the reconstructed functions behave similarly to the observational data, but with less propagation of errors.

\section{Discussion and Conclusions}
In this paper we presented observational constraints on some of the cosmographic functions, specially
the deceleration parameter, the jerk parameter and the snap parameter using a model-independent approximation. 
The statistical method used has been the Gaussian processes through the $GaPP$ public package.
The data used for this study are observational and simulated data from the Hubble parameter.
We directly estimate the value of the Hubble constant from the data used, 
$ H_{0} = 69.45 \pm 4.34 $, and use this value to perform our simulations.
However, in order to observe the effect of priors values of the Hubble constant on the functions
$q(z)$, $j(z)$ and $s(z)$ we use the values from the Planck collaboration and from the $SH0ES$ collaboration \cite{shoes}.

On the other hand, even though it is not our objective to discuss the Hubble tension, it is interesting to note that
our result marginally includes both the Planck result and the SH0ES result, see figure 6. However, this is primarily due to the large dispersion of the $CC$ measures. But what is interesting is to observe that following the same methodology presented here and using data from future observational projects such as the $LSST$ the $CC$ measurements can become excellent discriminators for the Hubble tension problem.

We propose to use the cosmographic functions given by the equations (15-17) instead of the 
cosmographic parameters derived from Taylor's expansion commonly used in the literature.
By using cosmographic functions we avoid the Taylor's expansion truncation problem
and consequently we can use data with $z> 1$ without restrictions. 
Also if we use cosmographic functions it is not necessary a new redshift variable 
 which can lead to inconsistent results. 

In general we can observe that the current observational data of the Hubble parameter do not allow to exclude the model $\Lambda CDM $ which still fits the reconstruction of the cosmographic functions. On the other hand, the simulated data, in particular, the Hubble parameter data obtained using the BAO technique seems promising to distinguish between alternative models to the $ \Lambda CDM $ model. 

An interesting application of our reconstruction is the study of specific dark energy models by writing the parameters of a given dark energy model as a function of the cosmographic parameters.

To implement this methodology we can proceed as follows: given a specific Hubble function we can use our system of equations (15-17) to determine a system of algebraic equations between the cosmographic parameters $ (q_{0}, j_{0} , s_{0}) $ and the parameters of a given model $ (\theta_ {i}) $. Then, we can solve the model parameters as a function of the cosmographic parameters: $ \theta_{i} = \theta_{i} (q_{0}, j_{0}, s_{0}) $. 

As an instructive example in the present work we are going to apply this methodology to a two-parameter cosmological model, 
specifically the $CPL$ model.
In this case the Hubble function is given by equation (5) so that the parameters of the model are $ \theta_{i} = (w_{0}, w_{1}, \Omega_{m0}) $, where initially  we assume the parameter $\Omega_{m0}$ as known. This means that to determine the parameters $(w_{0}, w_{1})$ we only need equations (15) and (16). Thus, we get an algebraic system of equations that provide us with the following solutions: 

\begin{eqnarray}
w_{0} &=&-\frac{1}{(1-\Omega_{m0})} + \frac{2(q_{0}+1)}{3(1-\Omega_{m0})},
\end{eqnarray}

\begin{eqnarray}
w_{1} &=& -\frac{2}{3}\frac{(j_{0}-1)}{(\Omega_{m0}-1)} -3w_{0} -3w_{0}^{2}.
\end{eqnarray}

If we consider the best fits and the value $\Omega_{m0} = 0.315$ (Planck) we obtain for the  parameters: $w_{0} =-1.132, w_ {1}=-0.461$.

Additionally, if we consider the equation for the snap parameter, equation (17), we can also determine the parameter $ \Omega_{m0} $ as a function explicitly of the cosmographic parameters so after some manipulations we obtain, 
\begin{align}
\Omega_{m0} &= \frac{-1 + 9 j_0 + 6 q_0 + 10 j_0 q_0 - 10 q_0^2 + 2 q_0^3}{2 (-2 q_0 + (3 + q_0) (2 j_0 + q_0^2) + s_0)}\\& \nonumber
+ \frac{(1 - 2 q_0)}{2 (-2 q_0 + (3 + q_0) (2 j_0 + q_0^2) + s_0)}\\& \nonumber
\times \left(1 + 9 j_0^2 + 2 j_0 (-9 - 4 q_0 + 8 q_0^2) + 
 4 q_0 (q_0 (9 + q_0 (5 + 2 q_0)) + 2 (-1 + s_0)) - 4 s_0 \right)^{1/2}\\& \nonumber
+\frac{s_{0}}{2 (-2 q_0 + (3 + q_0) (2 j_0 + q_0^2) + s_0)}.    \nonumber
\end{align}

Now we can use this equation to derive the value of $ \Omega_ {m0} $ directly from the cosmographic parameters and then we can use this value to determine the parameters $ (w_{0}, w_{1}) $ using the equations (29-30).
If we do this we get: $ w_{0} = - 1.036, w_{1} = - 0.119, \Omega_{m0} = 0.247 $. 

In principle we can use this methodology for other cosmological models. If we add more cosmographic functions, then we can include models with more free parameters. In a subsequent investigation we intend to extend this methodology to other models of dynamic dark energy.
Therefore, it is important to note that in this way we can infer the values of the free parameters of a given model using the cosmographic approximation without using the Taylor expansion.

It is important to mention  that recently the reference \cite{escamilla} have proposed a similar method but in another context. They determine the cosmographic parameters as function of the parameters of different dark energy models and use observational data to estimate the values of the parameters of dark energy and, with these results, the cosmographic parameters.

As we mentioned, our methodology is very useful to avoid using the Taylor expansion, 
but in compensation it depends on the data of the Hubble parameter, which are still a small sample when compared to supernovae Ia, for example. 
Furthermore we need to reconstruct derivatives of observational data.  The fact that a small noise in the measurement data can cause a large error in the derivatives is a difficult problem to analyze. For this reason, we hope to investigate in a later work the use of different statistical methods to reconstruct the derivatives as well as to discuss the importance of data quality.

\begin{figure*}
	\includegraphics[width=5.00in,height=3.00in]{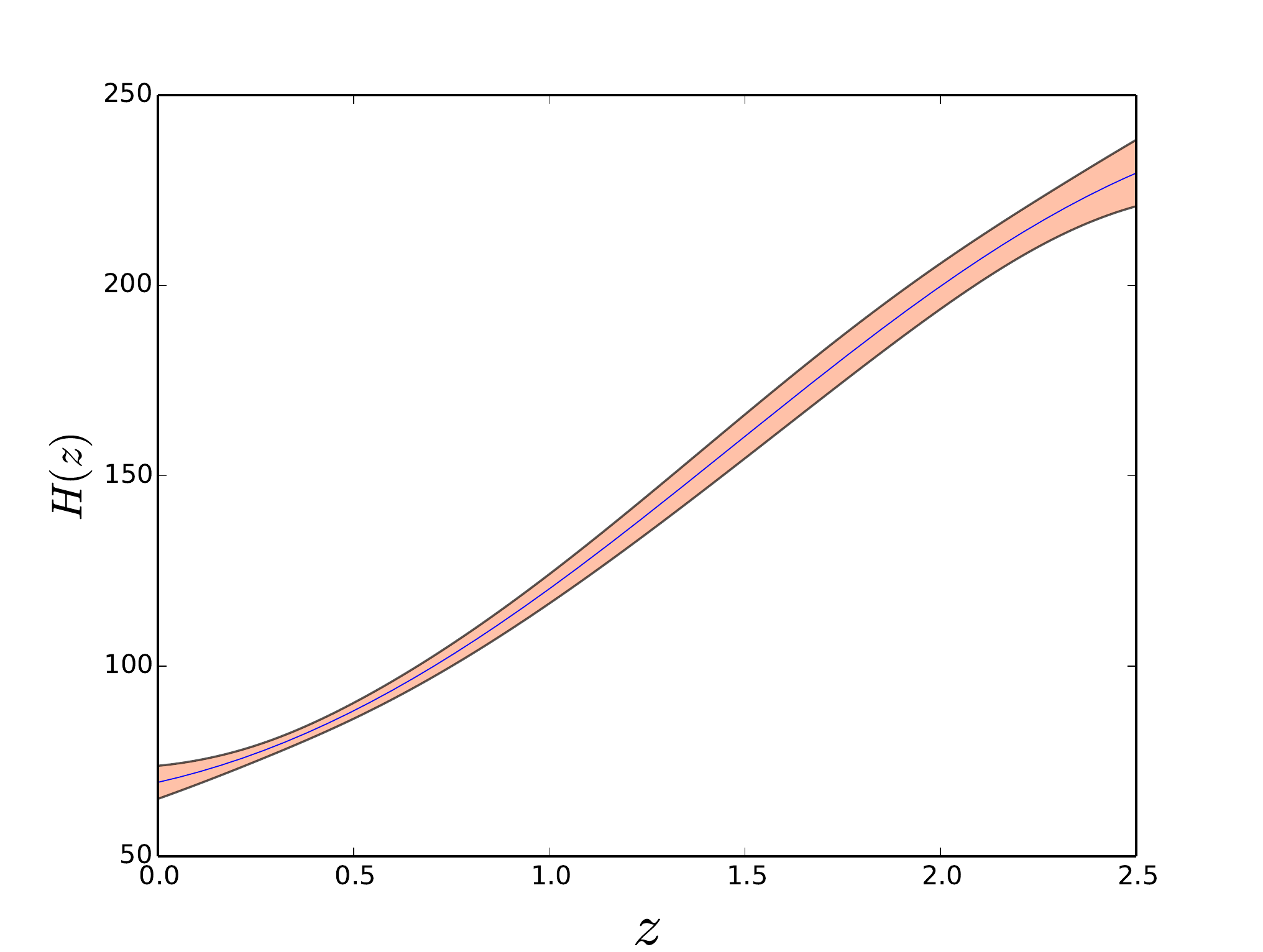}
	\includegraphics[width=5.00in,height=3.00in]{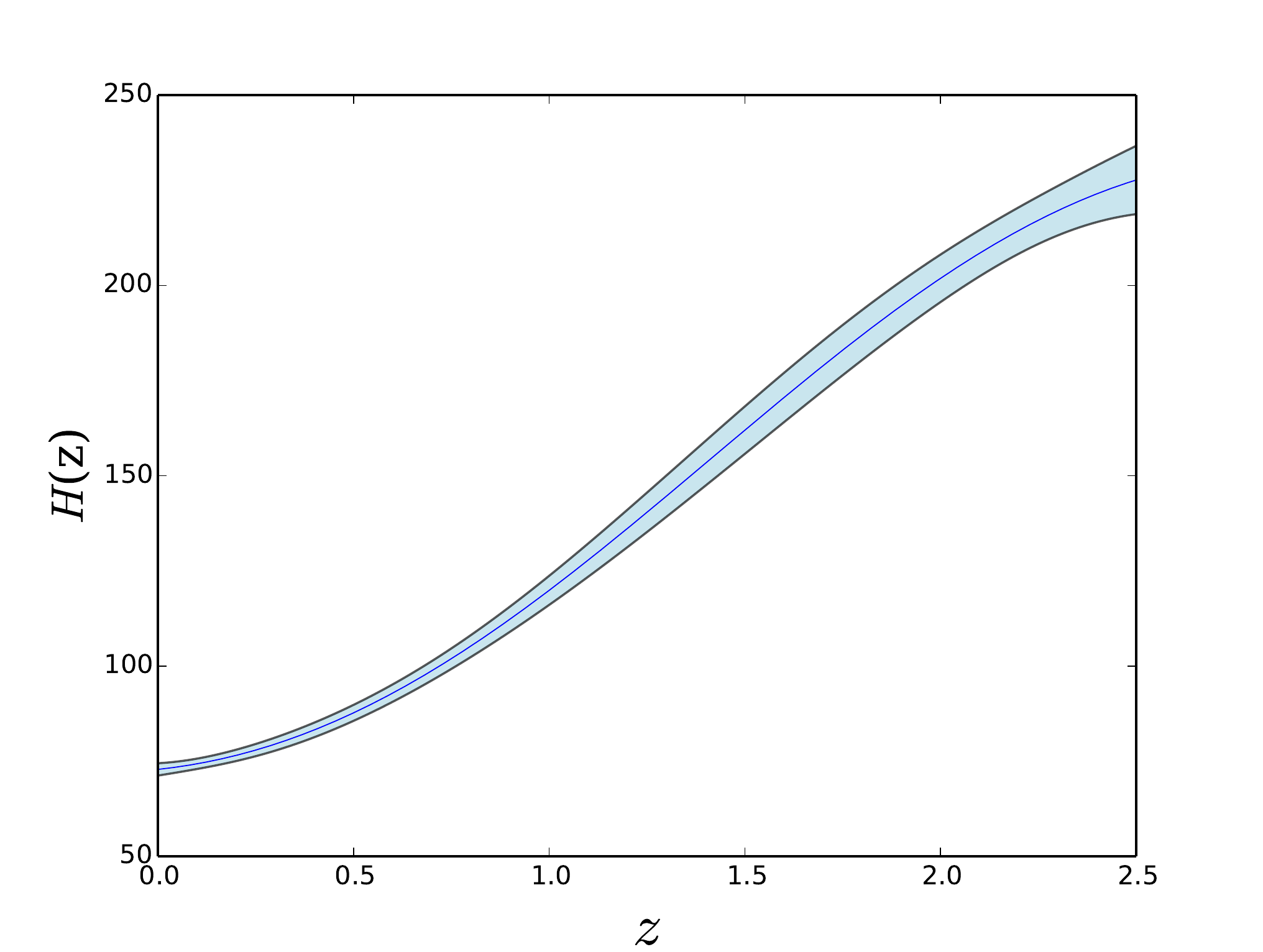}
	\includegraphics[width=5.00in,height=3.00in]{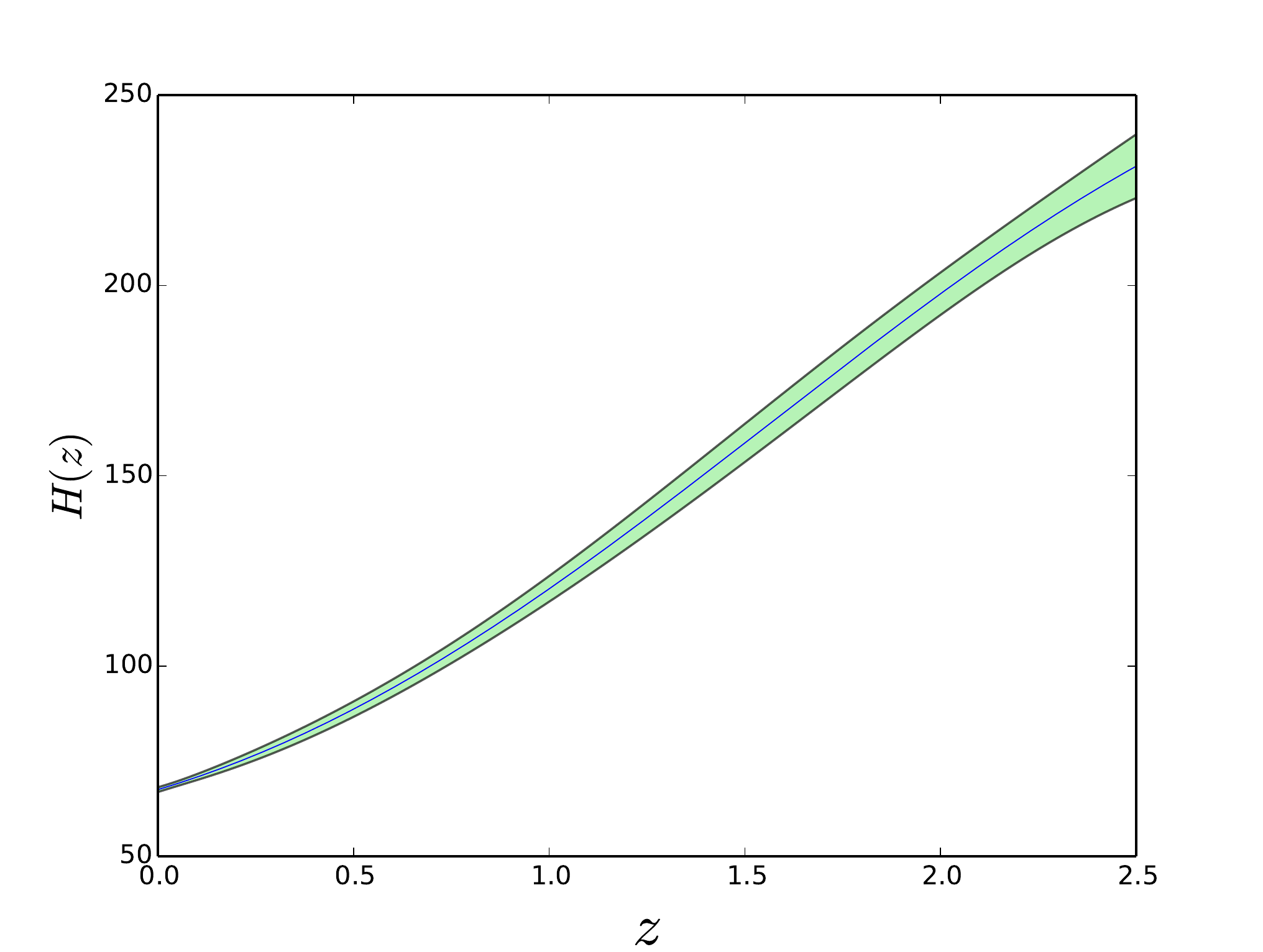}
    \caption{In the top the reconstruction of the Hubble Parameter using observational data
		of Hubble parameter in a model-independent form. In the middle the reconstruction of the Hubble Parameter with a prior $H_{0}=74.03$ ($SH0ES$ collaboration). 
    In the bottom the reconstruction using a prior $H_{0}=67.44$ (Planck's collaboration).}
\end{figure*}

\begin{figure*}
	\includegraphics[width=3.7in,height=2.5in]{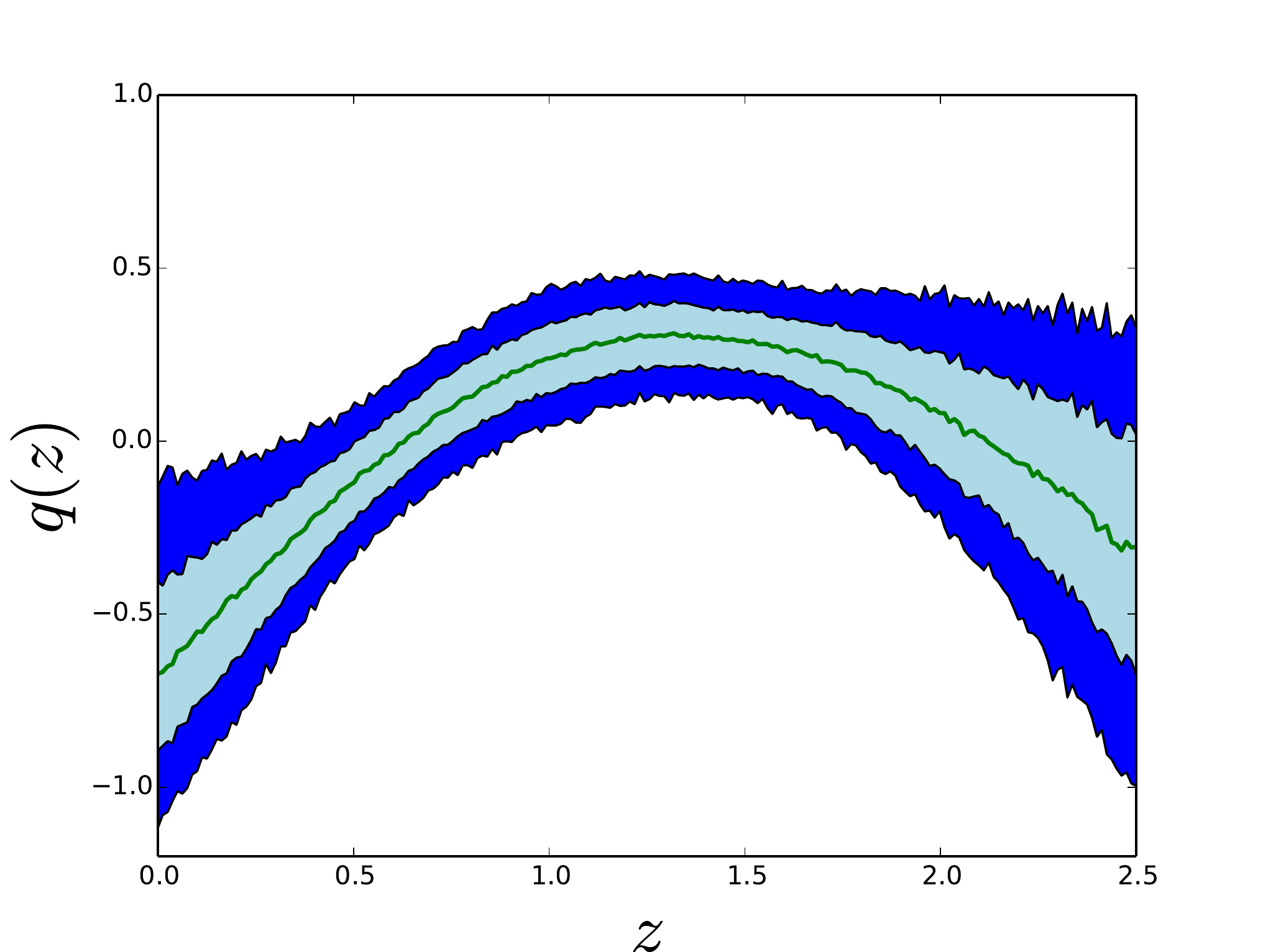}
	\includegraphics[width=3.7in,height=2.5in]{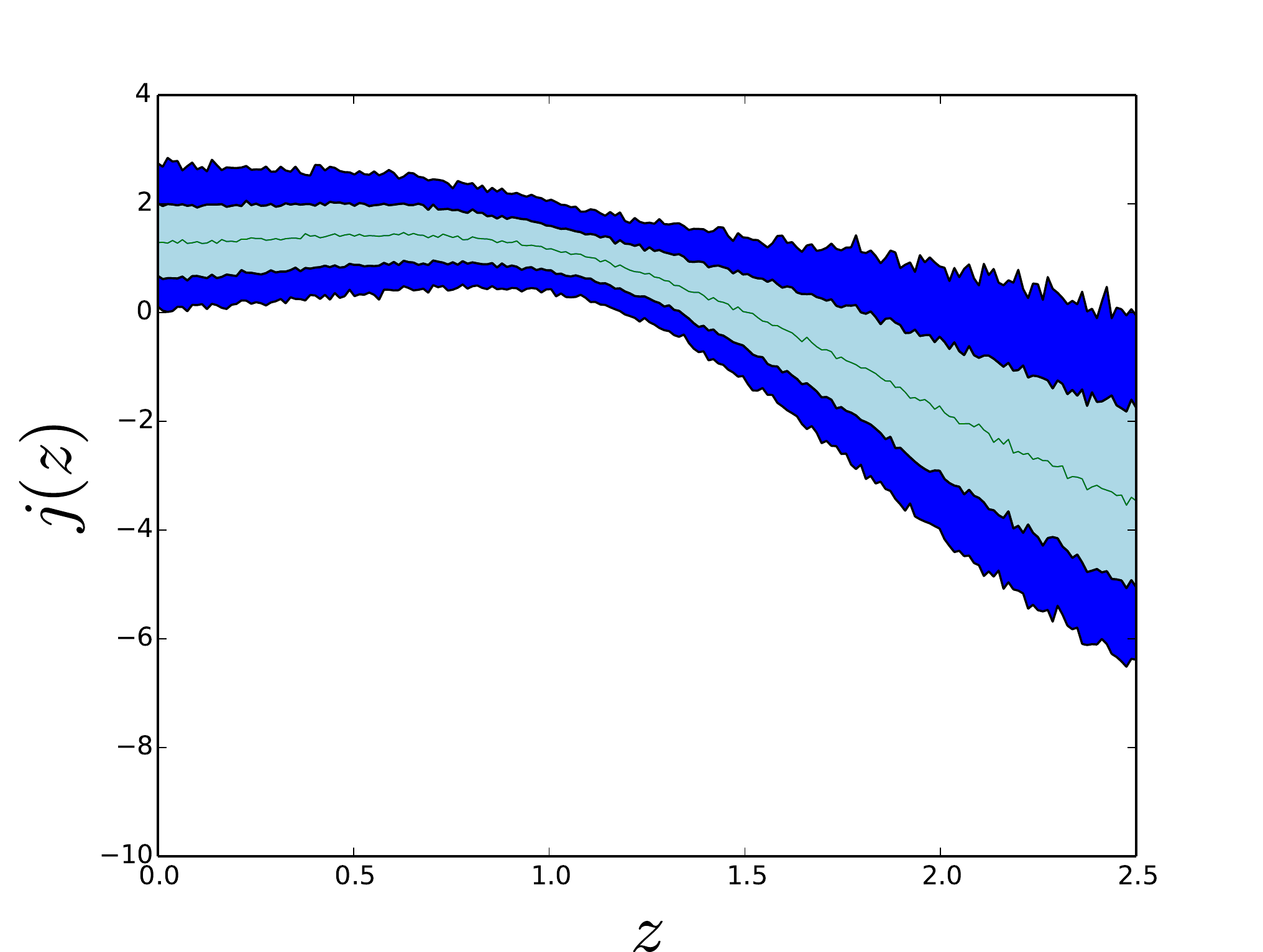}
	\includegraphics[width=3.7in,height=2.5in]{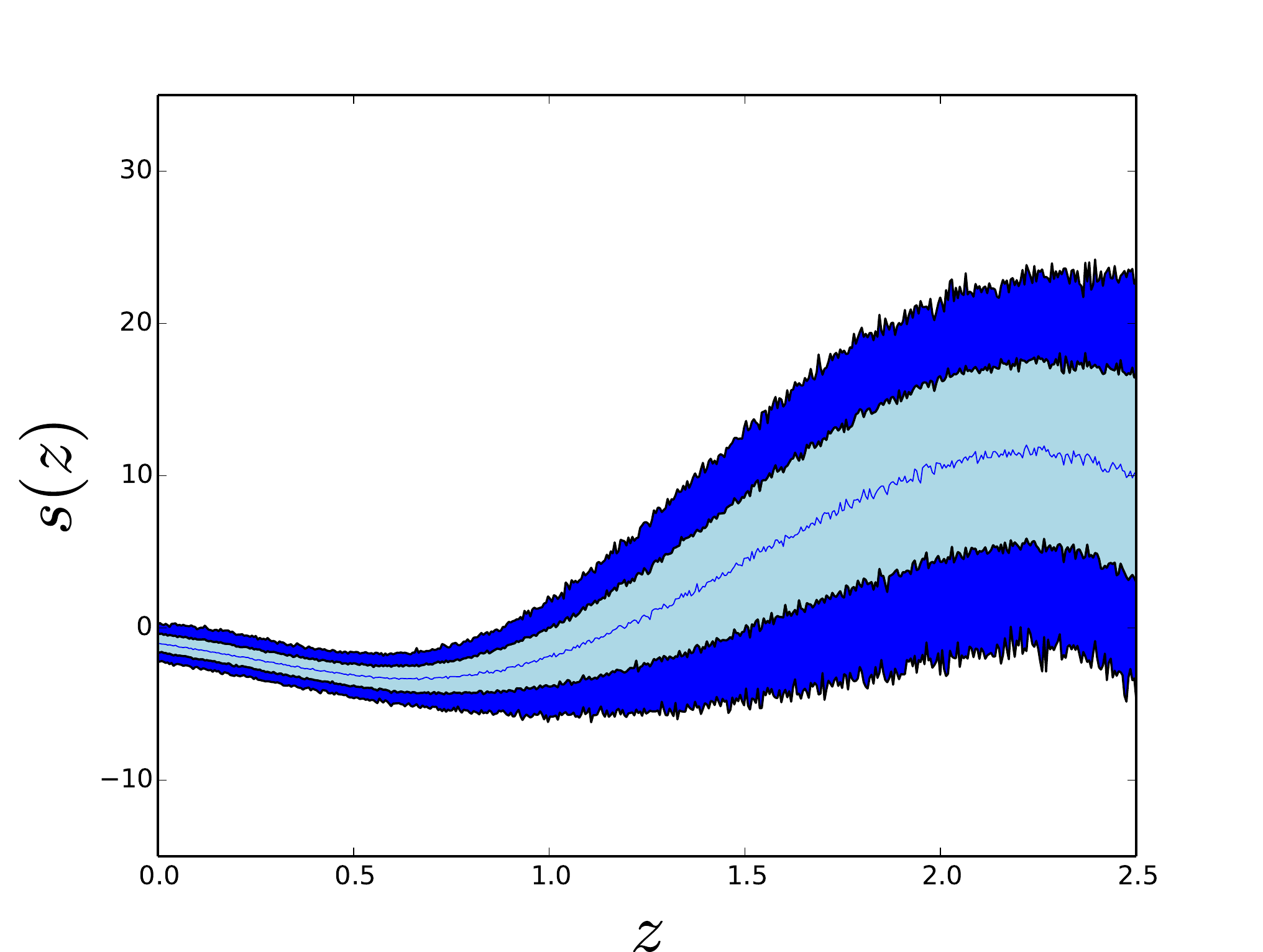}
    \caption{Observational constraints on the cosmography functions using the observational 
    data of the Hubble parameter in a model-independent form.}
    \label{fig:example*}
\end{figure*}

\begin{figure*}
	\includegraphics[width=5.5in,height=7.5in]{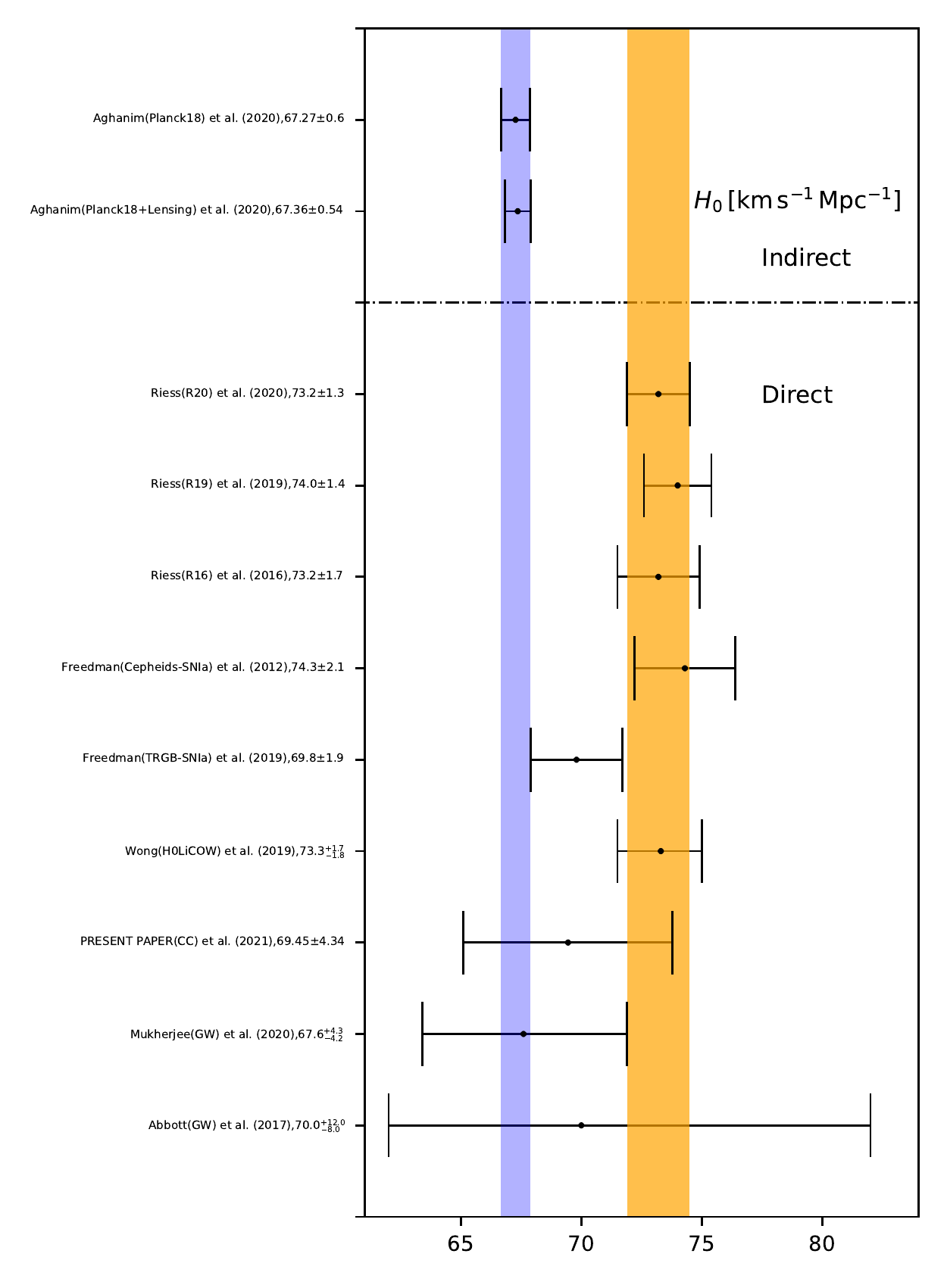}
    \caption{Observational constraints on the Hubble constant, $H_{0}$, where we show our result in comparison with the main results published in the literature, for example, compiled in \cite{divalentino}. The orange vertical band corresponds to the SH0ES team and the blue vertical band corresponds to the Planck Collaboration. The figure was made using the public program available online at github.com/lucavisinelli/H0TensionRealm. \cite{divalentino}.}
    \label{fig:example*}
\end{figure*}

The figure was made using the public program available at

\begin{figure*}
	
	\includegraphics[width=3.0in,height=2.5in]{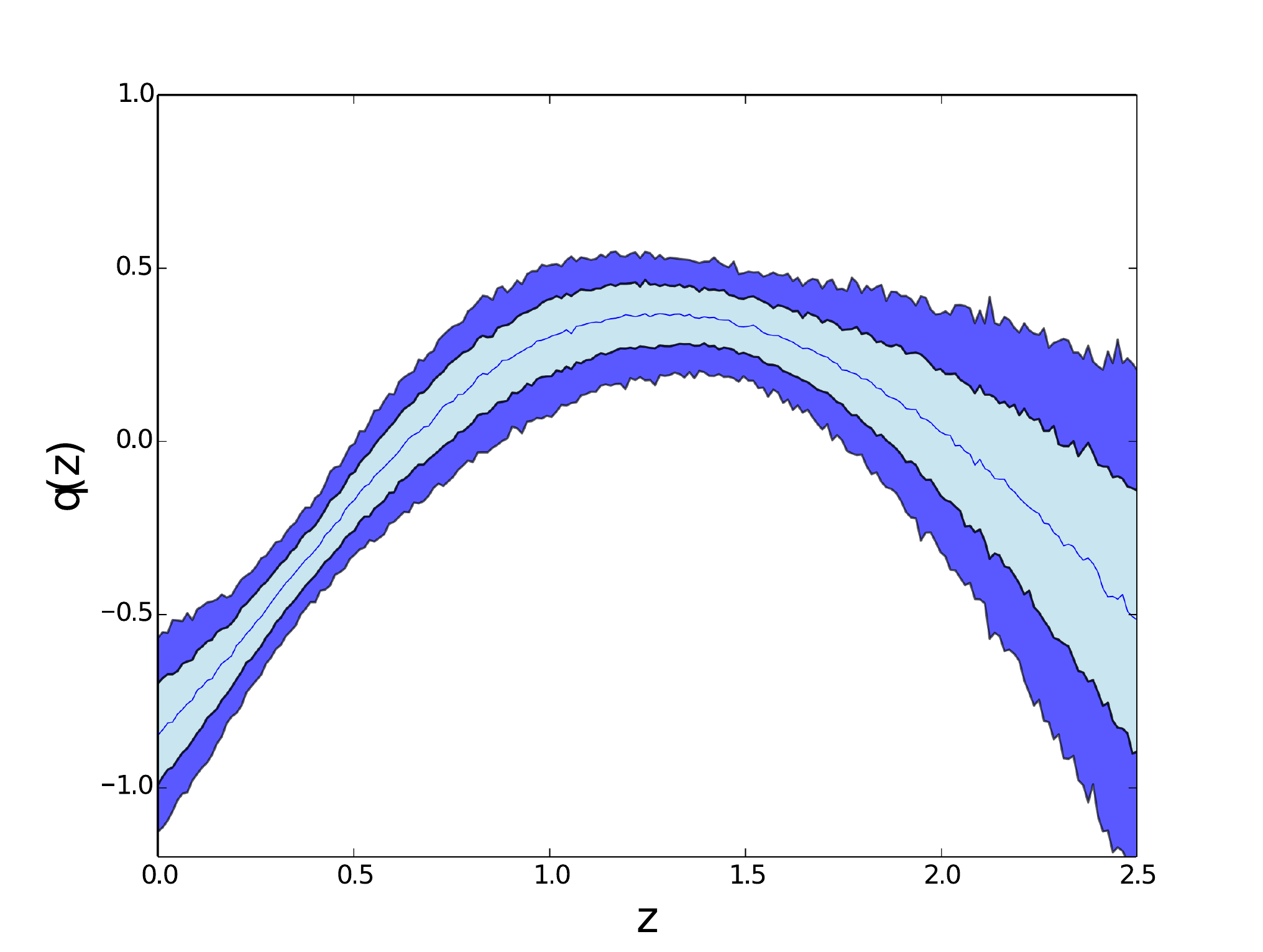}
	\includegraphics[width=3.0in,height=2.5in]{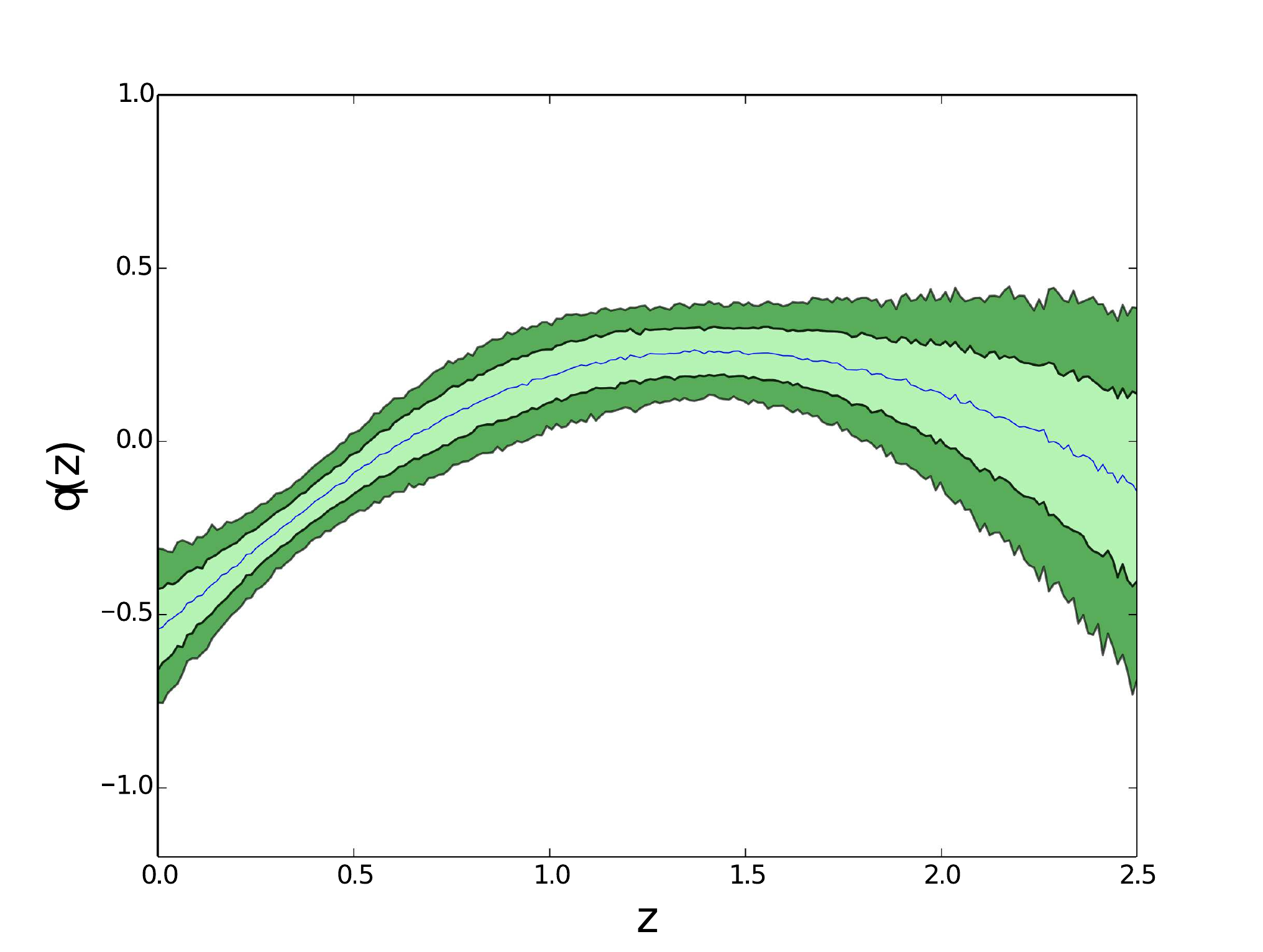}
	\includegraphics[width=3.0in,height=2.5in]{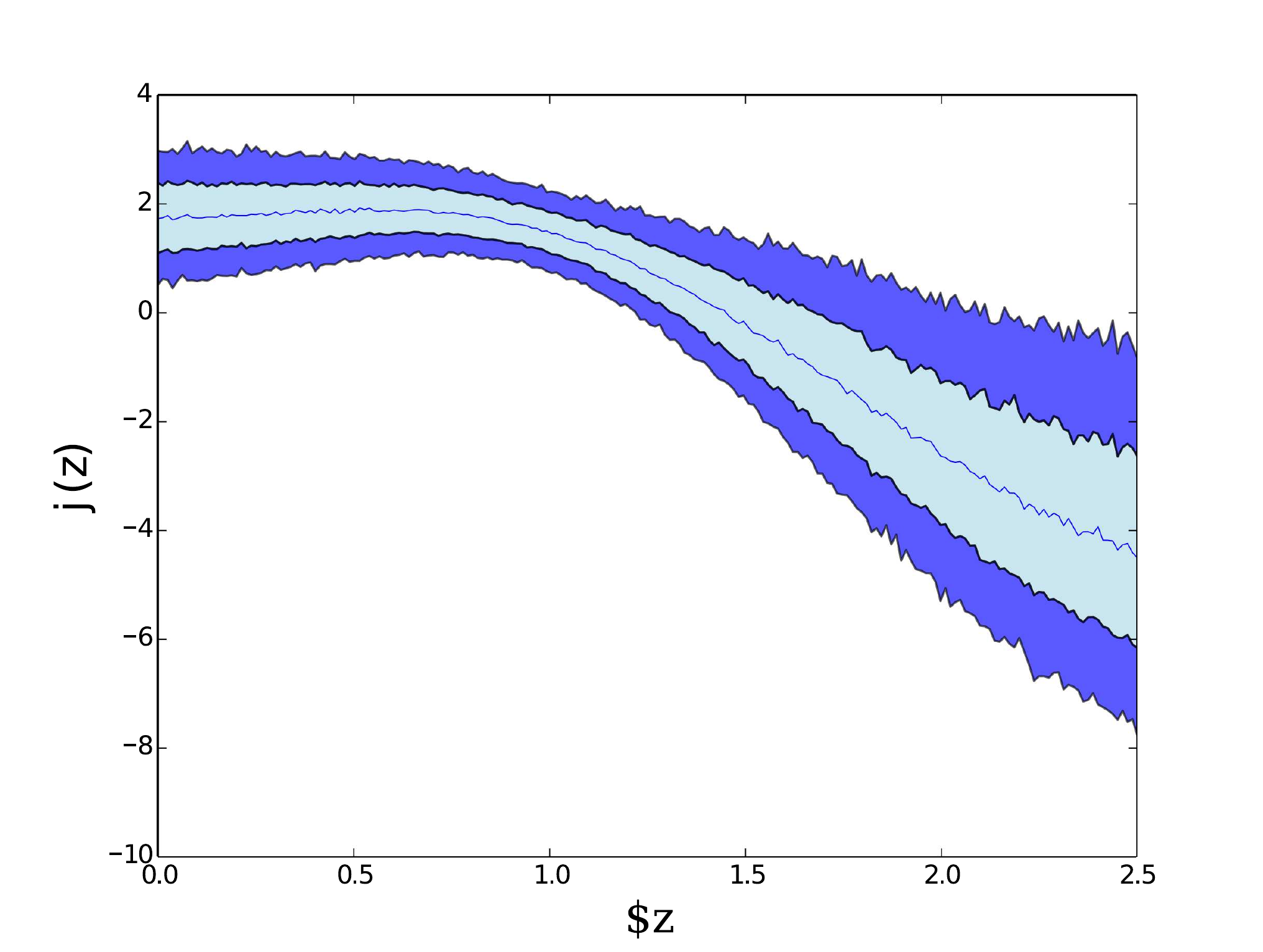}
	\includegraphics[width=3.0in,height=2.5in]{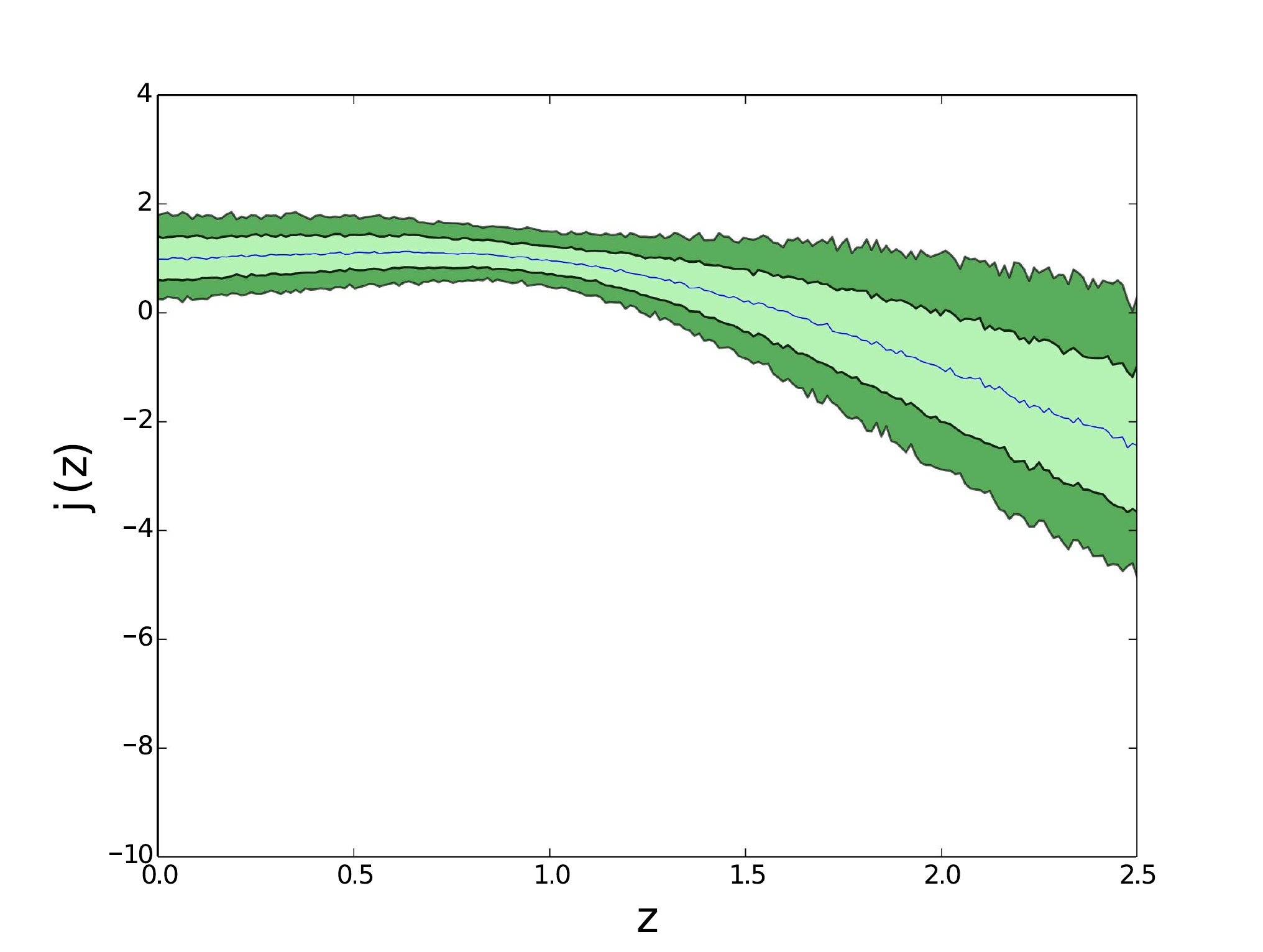}
	\includegraphics[width=3.0in,height=2.5in]{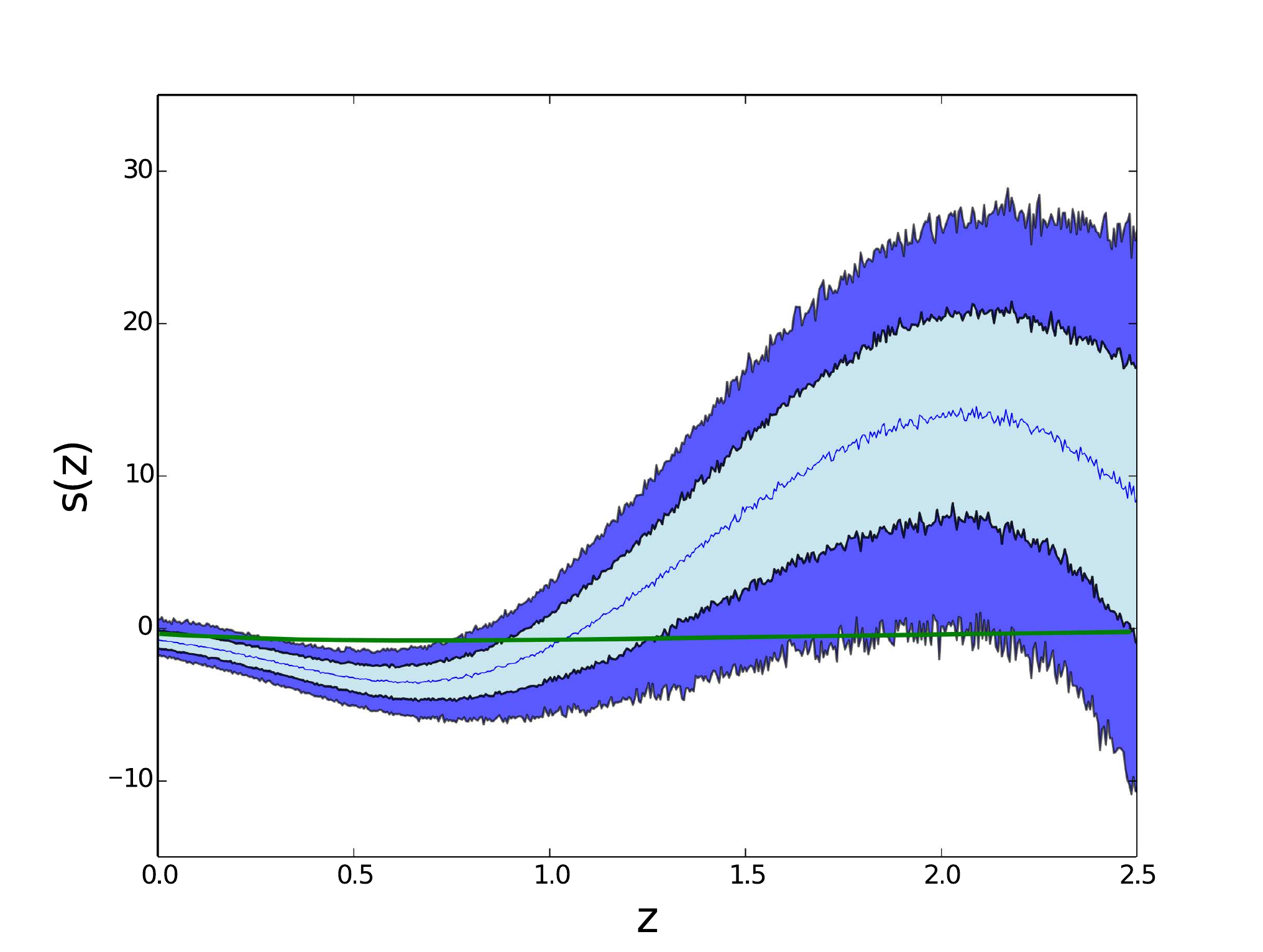}
	\includegraphics[width=3.0in,height=2.5in]{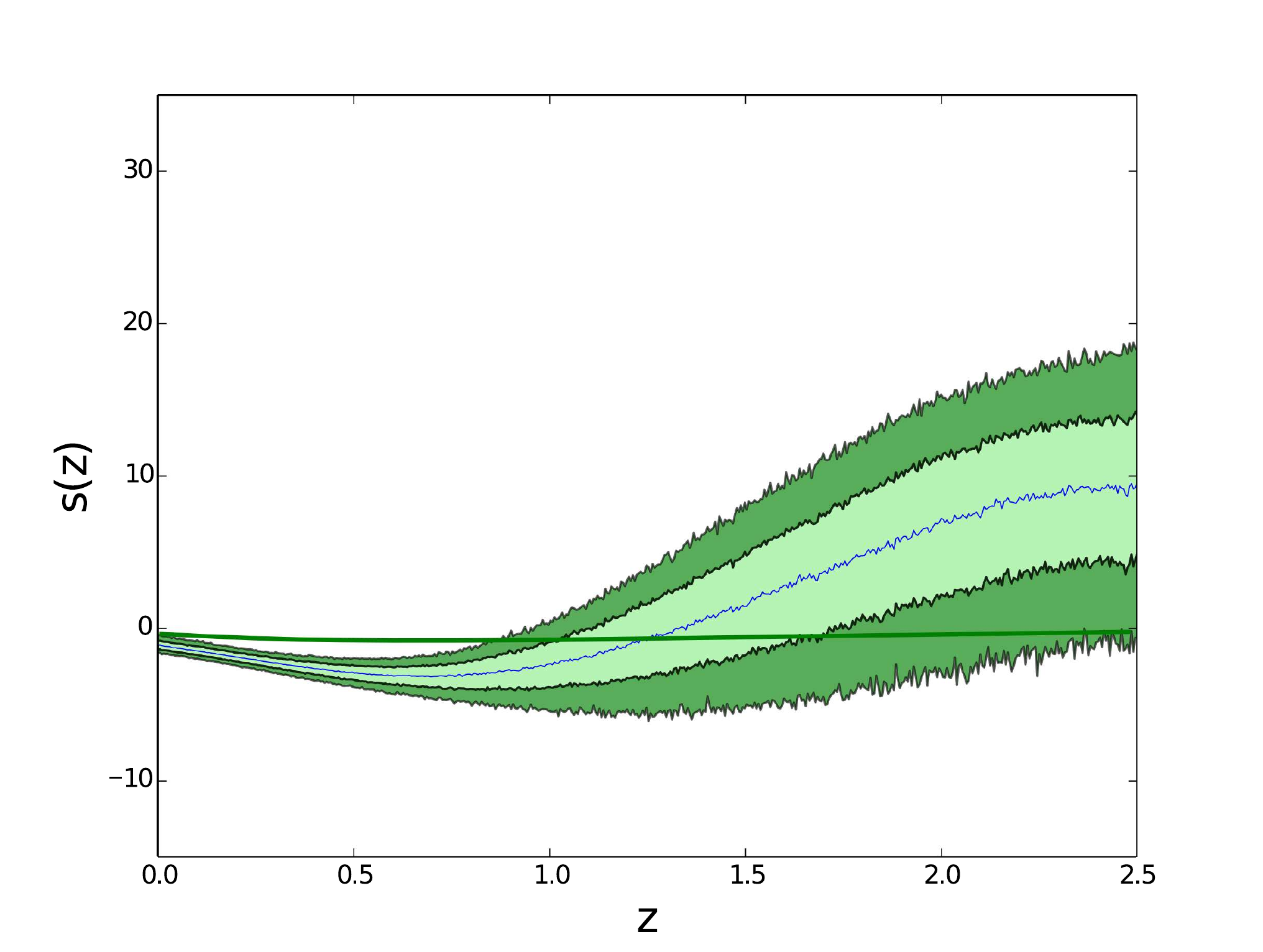}
    \caption{Observational constraints on the cosmography functions using the observational 
    data of Hubble parameter for the two values of the Hubble constant. 
		In left side for $H_{0} =  74.03$ and in right side for $H_{0} = 67.44$.}
    \label{fig:example*}
\end{figure*}

\begin{figure*}
	\includegraphics[width=3.0in,height=2.5in]{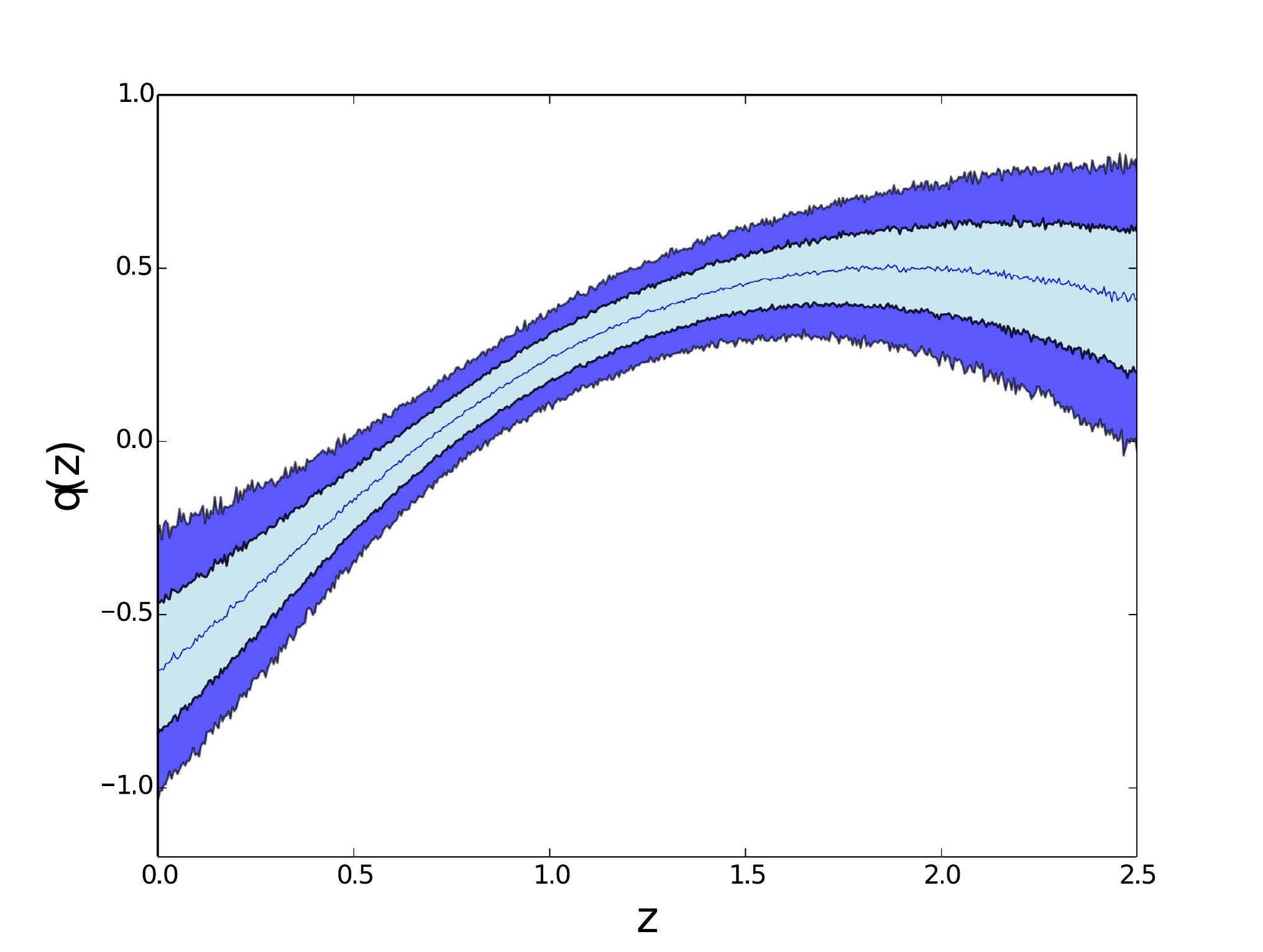}
	\includegraphics[width=3.0in,height=2.5in]{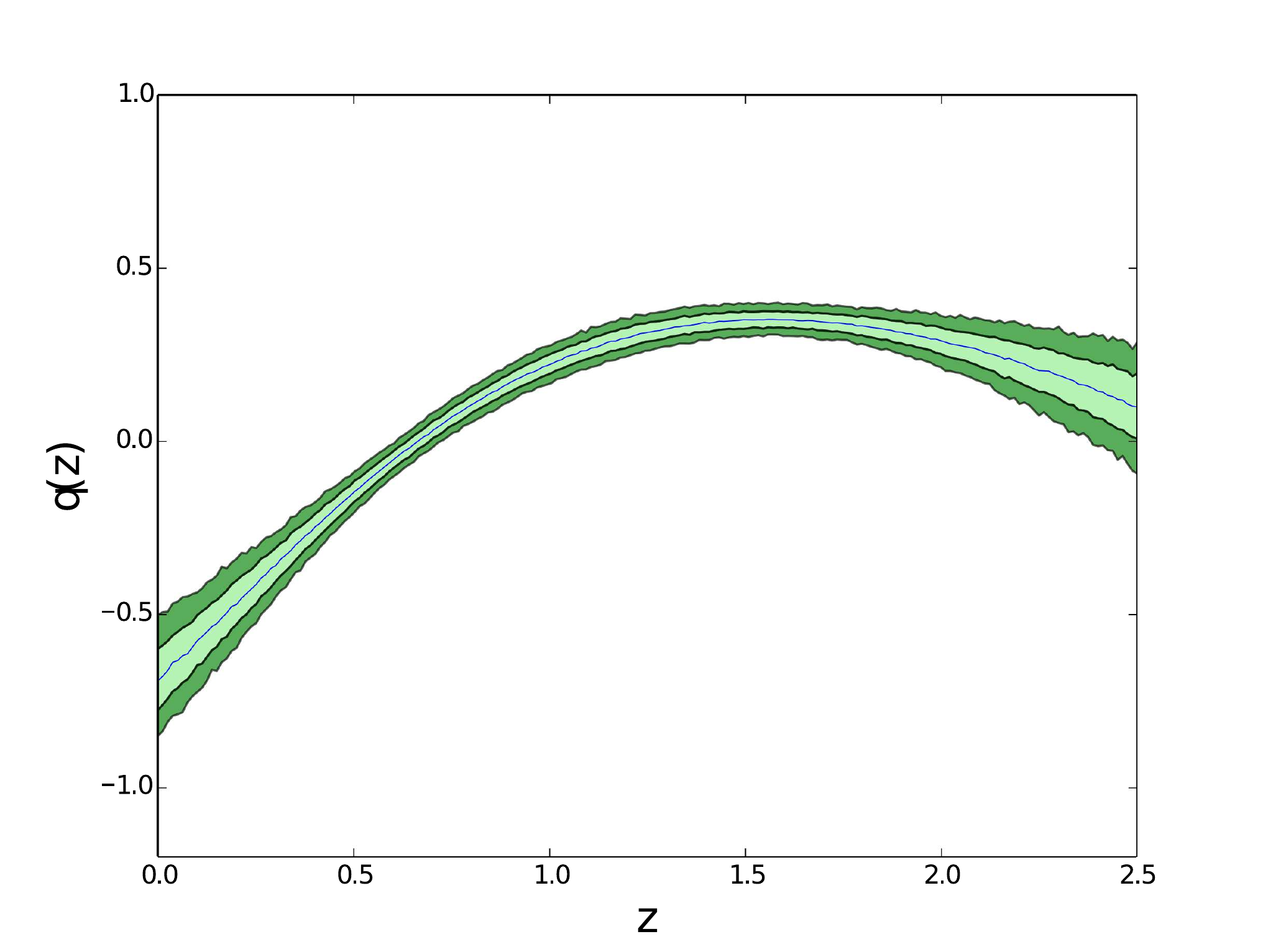}
	\includegraphics[width=3.0in,height=2.5in]{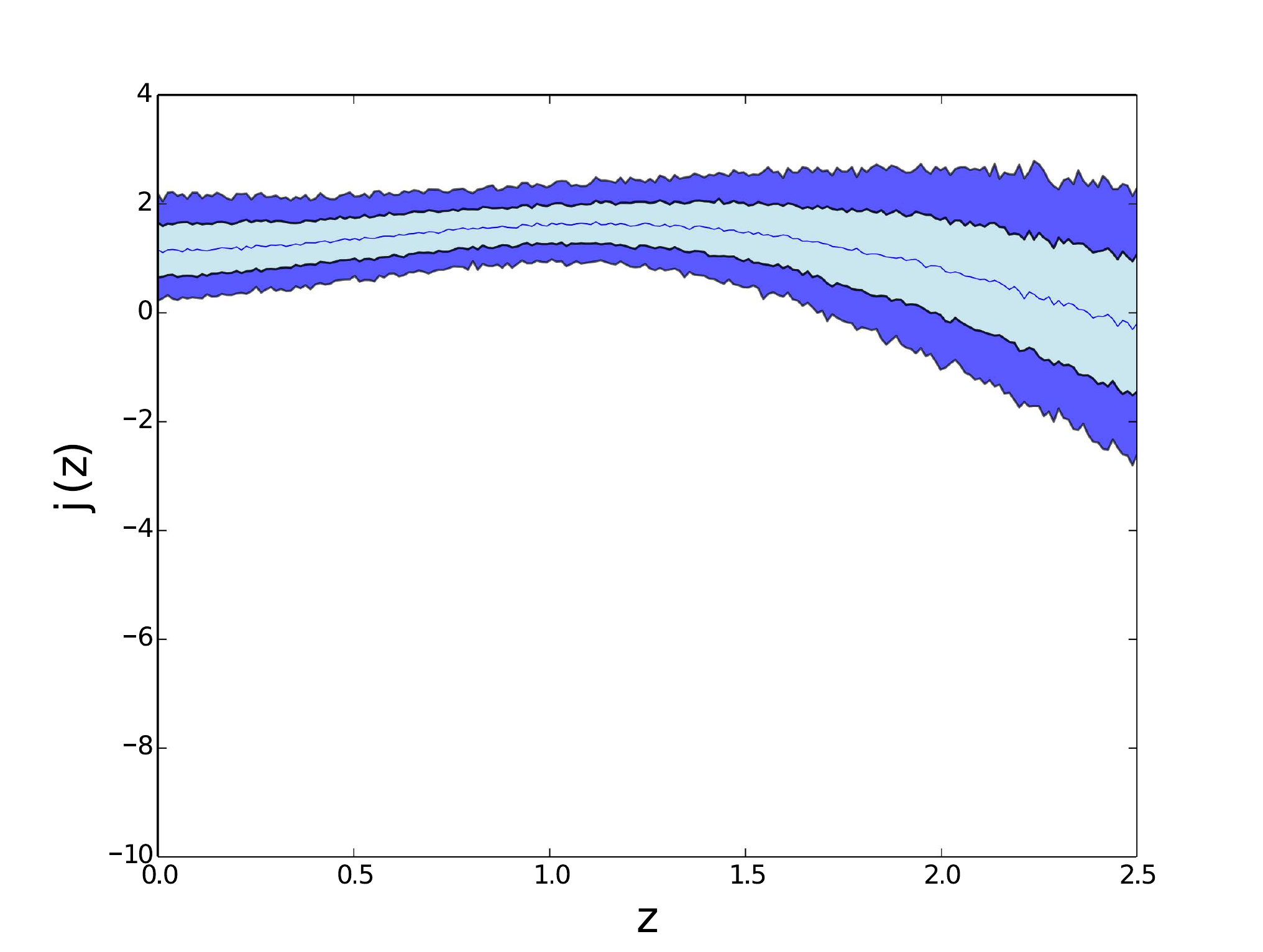}	
	\includegraphics[width=3.0in,height=2.5in]{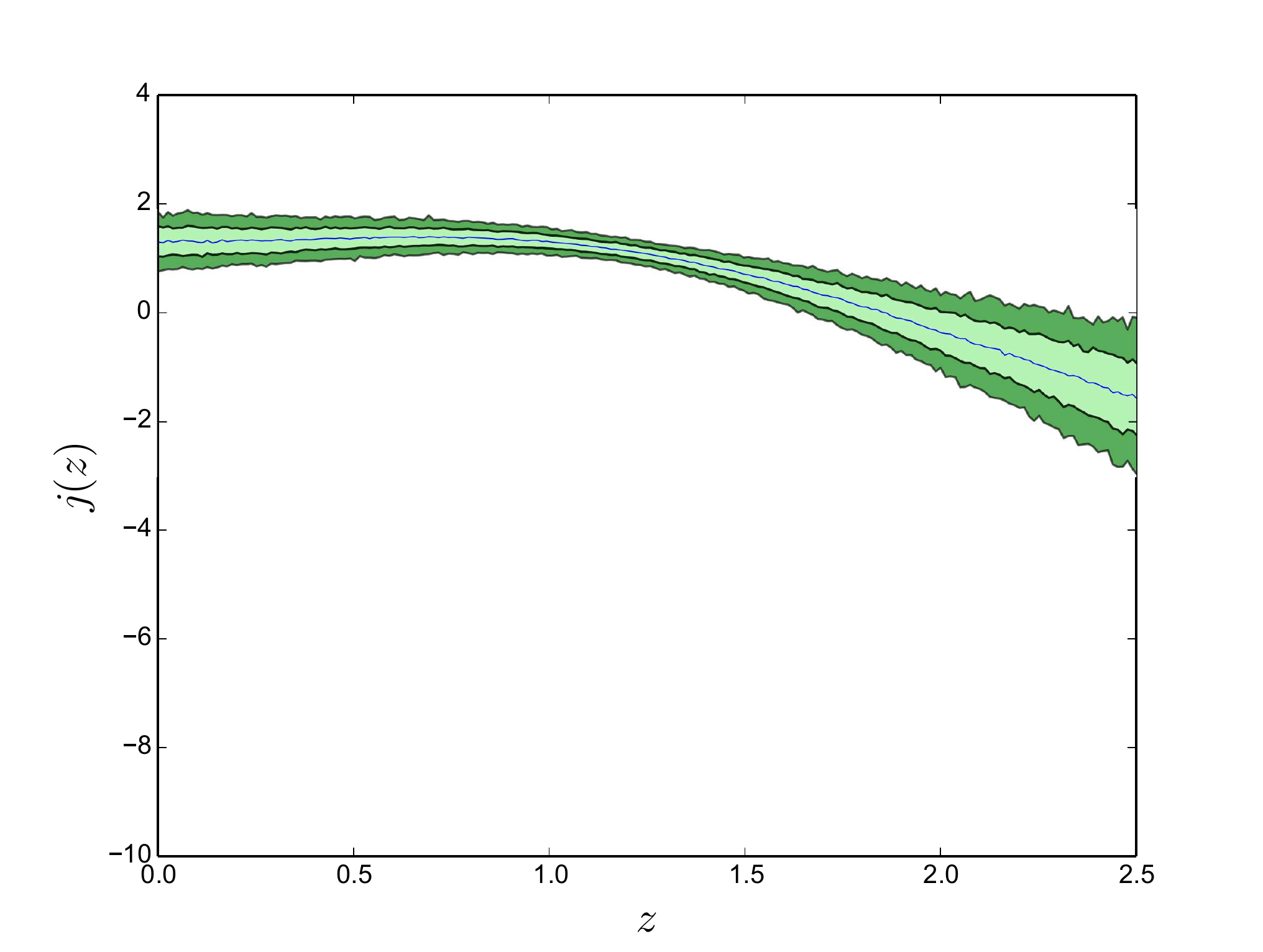}
	\includegraphics[width=3.0in,height=2.5in]{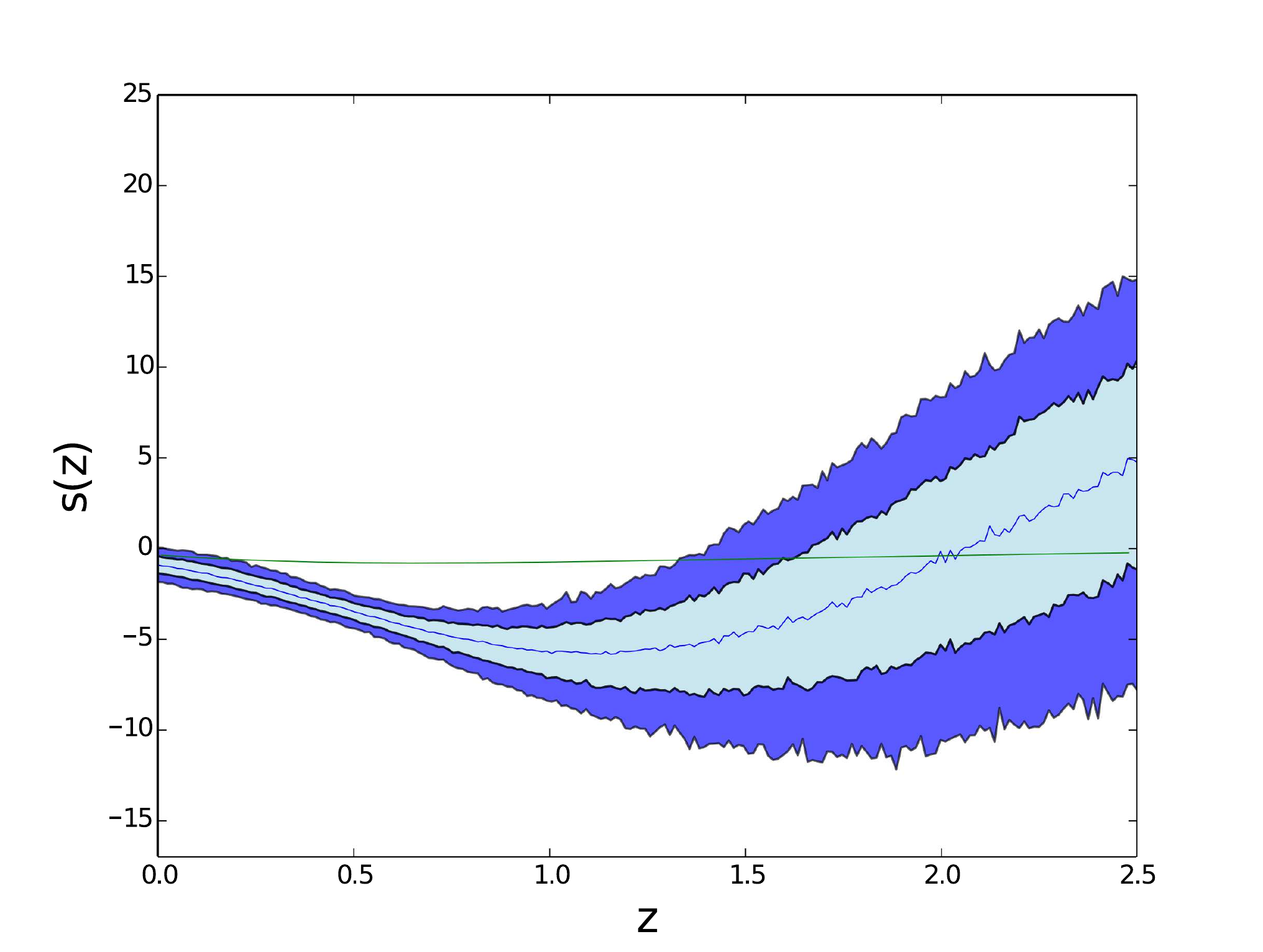}
	\includegraphics[width=3.0in,height=2.5in]{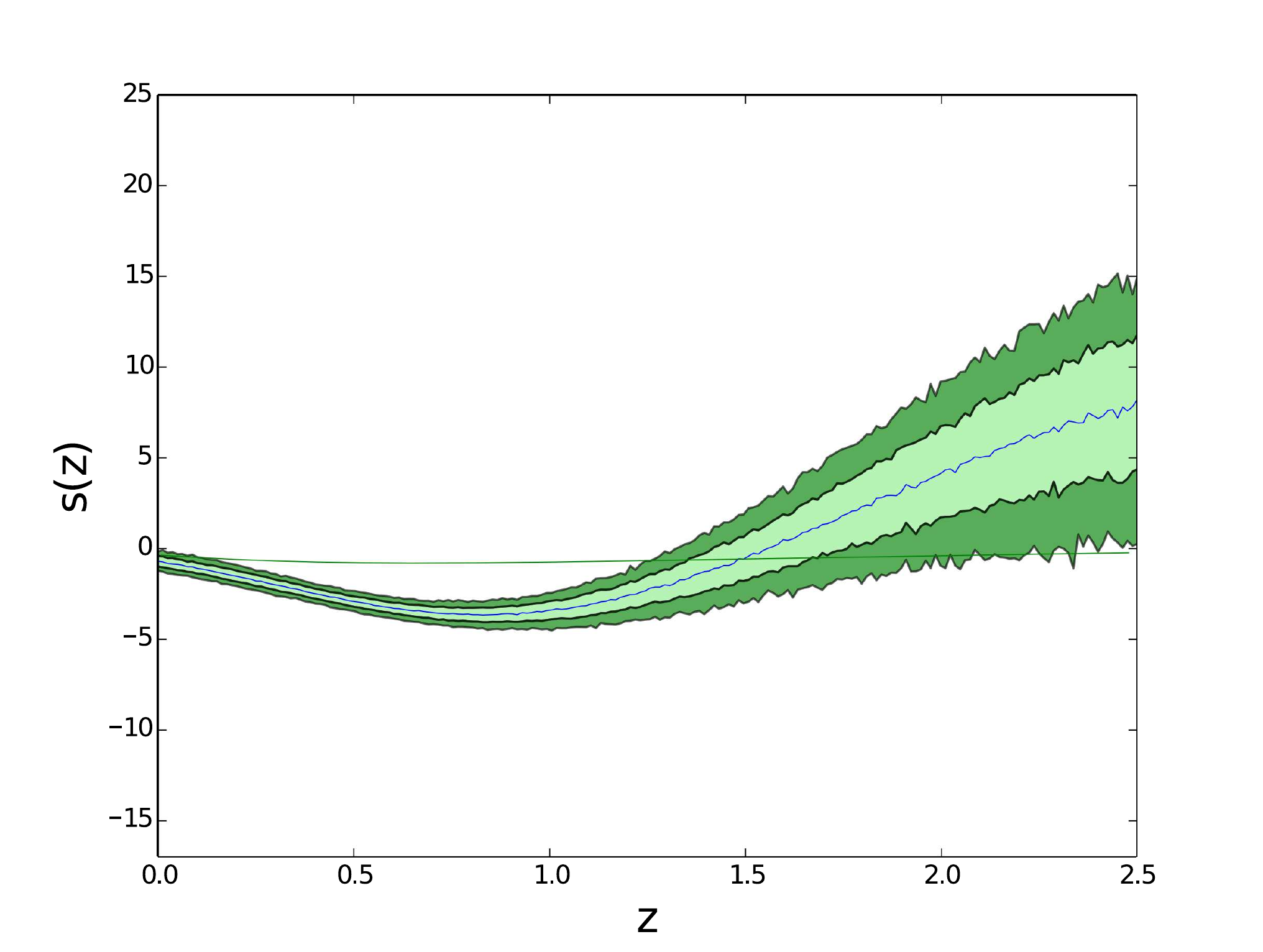}
        \caption{In the column on the left we used the simulation that uses the $CC$ prescription and in the 
        right column we have used the $BAO$ prescription. In both cases we have used the value of the Hubble 
        parameter estimated directly from the observational data. The value used is in Table 2. }
\end{figure*}

\begin{table*}
\caption {The best-fitting parameter values and uncertainties with $1\sigma$ error bars for the cosmographic parameters
from the observational and simulated data. The value $69.45$ $\pm$ $4.34$ is the best-fit for all observational data (CC + BAO) 
and is assumed as input for our simulations.} \label{tab:title} 
\centering
\begin{tabular}{ccccc}
\\ \hline
\hline
Data               & $H_{0}$($1\sigma$)                 & $q_{0}$($1\sigma$)            & $j_{0}$($1\sigma$)          & $s_{0}$($1\sigma$)                      
\\ \hline
\hline
CC+BAO &    $69.45$ $\pm$ $4.34$    & $-0.672^{+0.265}_{-0.223}$ & $1.200^{+0.300}_{-0.390}$ & $-0.670^{+0.250}_{-0.390}$
\\
CC+BAO &  $67.44$ $\pm$ $0.58$ (Value of Planck)    & $-0.670^{+0.210}_{-0.120}$ & $0.985^{+0.414}_{-0.380}$ & $-0.600^{+0.379}_{-0.300}$
\\
CC+BAO  & $74.03$  $\pm$ $1.42$ (Value of SH0ES)   & $-0.710^{+0.159}_{-0.111}$ & $1.714^{+0.593}_{-0.592}$ & $-0.700^{+0.670}_{-0.567}$
\\
Simulation CC      & $69.45$ $\pm$ $4.34$              & $-0.657^{+0.123}_{-0.187}$ & $1.155^{+0.501}_{-0.509}$ & $-0.899^{+0.485}_{-0.458}$
\\
Simulation BAO      & $69.45$ $\pm$ $4.34$             & $-0.690^{+0.092}_{-0.085}$  & $1.291^{+0.285}_{-0.263}$ & $-0.716^{+0.294}_{-0.285}$
\\ \hline \hline
\end{tabular}
\end{table*}

\section*{Acknowledgements}
A.M.V.T appreciate the computational facilities of the UFES to develop the present investigation and also the authors thank Célia Escamilla-Rivera for helpful discussions on the interpretation of cosmographic parameters. J. C. Fabris thanks Fundação de Amparo
Pesquisa e Inovação do Espirito Santo (FAPES, project number 80598935/17) and Conselho Nacional de Desenvolvimento
Científico e Tecnológico (CNPq, grant number 304521/2015-9) for partial support.

\appendix

\section{Supernovae Ia}
In this study we use the data from Supernovas Ia called "Pantheon" sample (\cite{scolnic}) which is the largest
combined sample of SNIa and consists of $1048$ data with redshifts in the range $0.01 < z < 2.3$.
 It is a collection of SNe Ia discovered by the Pan-STARRS1 (PS1) Medium Deep Survey
and SNe Ia from Low-z, SDSS, SNLS and HST surveys.
This supernova Ia compilation uses The SALT 2 program to transform light curves into distances using a modified version of the Tripp 
formula \cite{tripp},
\begin{equation}
\mu = m_{B} - M + \alpha x_{1}-\beta c + \Delta_{M} + \Delta_{B},
\end{equation}
where $\mu$ is the distance modulus, $\Delta_{M}$ is a distance correction based on the host-galaxy mass of the SNIa and $\Delta_{B}$
is the distance correction based on predicted bias from simulations. Also $\alpha$ is the coefficient of the relation between luminosity and stretch, $\beta$ is the coefficient of the relation between luminosity and color and $M$ is
the absolute $B$-band magnitude of a fiducial SNIa with $x_{1} = 0$ and $c = 0$. Also $c$ is the color and $x_{1}$ is the light-curve shape parameter and $m_{B}$ is the log of the overall flux normalization.
An uncertainty matrix $\bf{C}$ is defined such that,
\begin{equation}
\chi_{SNIa}^{2} = \Delta \vec{\mu}^{T}. \mathbf{C}^{-1}. \Delta \vec{\mu},
\end{equation}
where $\Delta \vec{\mu} = \vec{\mu}_{obs}- \vec{\mu}_{model}$ and $\vec{\mu}_{model}$ is a vector of distance modulus from a given cosmological model and $\vec{\mu}_{obs}$ is a vector of observational distance modulus. The distance module is defined as $\vec{\mu}=\vec{m}-M$, where $M$ is the absolute magnitude and $\vec{m}$ is the apparent magnitude which is given by the expression
\begin{align}
\vec{m}_{model} &= M+5Log_{10}(D_{L}) + 5Log_{10}(\frac{c/H_{0}}{1Mpc}) + 25 \\&
= \bar{M}+25+5Log(D_{L}).
\end{align}
where $D_{L}=\frac{H_{0}}{c}d_{L}$ and $\bar{M} = M+5Log(\frac{c/H_{0}}{1Mpc})$ is a nuisance parameter, which depends on the Hubble constant $H_{0}$ and the absolute magnitude $M$. To minimize with respect to the nuisance parameter we follow a process similar at the references 
\cite{conley} and \cite{arjona}. Therefore the $\chi^{2}_{\bar{M} marg}$ is,
\begin{equation}
\chi^{2}_{\bar{M} marg} = a+ \log \frac{e}{2\pi} -\frac{b^{2}}{e},
\end{equation}
where,
\begin{eqnarray}
a &=&  \Delta \vec{m}^{T} \cdot C^{-1}\cdot \Delta \vec{m},\\
b &=&  \Delta \vec{m}^{T} \cdot C^{-1} \cdot \mathrm{I},\\
e &=&  \mathrm{I}^{T} . C^{-1}. \mathrm{I}
\end{eqnarray}
where $\Delta \vec{m}= \vec{m}_{obs}-\vec{m}_{model}$ and $\mathrm{I}$ is the identity matrix.

\section{The Gaussianity of the data}
An important test of Gasussianity is to determine the parameter $N _ {\sigma}$ as shown in table 1.  
Then the weighted mean for the Hubble parameter is given by \cite{podariu},

\begin{equation}
 \bar{H} = \frac{\sum_{i=1}^{N}H_{i}(z_{i})/\sigma_{i}^{2}}{\sum_{i=1}^{N} 1/\sigma_{i}^{2}}.
\end{equation}
We also compute the weighted bin redshift, it is given by,

\begin{equation}
 \bar{z} = \frac{\sum_{i=1}^{N} z_{i}/\sigma_{i}^{2}}{\sum_{i=1}^{N} 1/\sigma_{i}^{2}},
\end{equation}
The associated weighted error is given by

\begin{equation}
\bar{\sigma} = (\sum_{i=1}^{N} 1/\sigma_{i}^{2})^{-1/2}.
\end{equation}
Using the above definitions we determine a reduced goodness-of-fit $ \chi^{2}$ for each bin as
\begin{equation}
 \chi_{\nu}^{2} = \frac{1}{N-1}\sum{\frac{H-\bar{H}}{\sigma_{i}^{2}}}.
\end{equation}
And thus the number of standard deviations with respect from unity is defined as
\begin{equation}
 N_{\sigma} = \mid \chi_{\nu}-1 \mid \sqrt{2(N-1)}.
\end{equation}

\end{document}